\documentclass[twocolumn,prb,superscriptaddress,showpacs]{revtex4}

\usepackage{amsmath,amssymb}
\usepackage{graphicx}
\usepackage{subfigure}
\usepackage{verbatim}
\usepackage{hyperref}
\usepackage{epstopdf}
\usepackage{amsfonts}
\usepackage{dcolumn}
\usepackage{bm}

\begin{document}

\title{New gapped quantum phases for $S=1$ spin chain with $D_{2h}$ symmetry}

\author{Zheng-Xin Liu}
\affiliation{Institute for Advanced Study, Tsinghua University,
Beijing, 100084, P. R. China}

\author{Min Liu}
\affiliation{Institute for Advanced Study, Tsinghua University,
Beijing, 100084, P. R. China}

\author{Xiao-Gang Wen}
\affiliation{Department of Physics, Massachusetts Institute of
Technology, Cambridge, Massachusetts 02139, USA}
\affiliation{Institute for Advanced Study, Tsinghua University,
Beijing, 100084, P. R. China}

\begin{abstract}
We study different quantum phases in integer spin systems with
on-site $D_{2h}=D_2\otimes Z_2$ and translation symmetry. We find
four distinct non-trivial phases in $S=1$ spin chains despite they
all have the same symmetry.  All the four phases have gapped bulk
excitations, doubly-degenerate end states and the doubly-degenerate
entanglement spectrum.  These non-trivial phases are examples of
symmetry protected topological (SPT) phases introduced by Gu and
Wen.  One of the SPT phase correspond to the Haldane phase and the
other three are new. These four SPT phases can be distinguished
experimentally by their different response of the end states to weak
external magnetic fields.  According to Chen-Gu-Wen classification,
the $D_{2h}$ symmetric spin chain can have totally 64 SPT phases
that do not break the symmetry.  Here we constructed seven
nontrivial phases from the seven classes of nontrivial projective
representations of $D_{2h}$ group.  Four of these are found in $S=1$
spin chains and studied in this paper, and the other three may be
realized in $S=1$ spin ladders or $S=2$ models.

\end{abstract}

\maketitle

\section{Introduction}

Topological order was introduced to distinguish different phases which can not
be separated by symmetry breaking orders.\cite{Wen90} Using a definition of
phase and phase transition based on local unitary transformations,
Ref.~\onlinecite{CGW1} shows that what topological order really describes
is actually the pattern of long range entanglements in gapped quantum systems.

The Haldane phase\cite{Haldanephase} for $S=1$ spin chains was regarded as a
simple example of topological order.  For a long time, the existence of string
order (or hidden $Z_2\otimes Z_2$ symmetry breaking), nearly degenerate end
states and gapped excitations were considered as the hallmark of the Haldane
phase.\cite{string} However, it was shown that even after we break the spin
rotation symmetry which destroy the string order and gapped end states, the
Haldane phase can still exist (i.e. is still distinct from the trivial phase).
It was also shown that the Haldane phase has no long range
entanglements.\cite{GuWen,CGW2} In fact, all 1D gapped ground state has no long
range entanglements.\cite{VCL0501}  Thus there are no intrinsic topologically
ordered states in gapped 1D systems\cite{CGW2} (including the Haldane phase).

But when Hamiltonians have some symmetries, even short range
entangled states with the same symmetry can belong to different
phases.\cite{GuWen,CGW1} Such phases are called `symmetry protected
topological (SPT) phases' by Gu and Wen.\cite{GuWen}  We would like
to remark that, due to their trivial intrinsic topological order and short
range entanglements, `symmetry protected topological phases' should
be more properly referred as `symmetry protected short range
entangled phases'. 

Thus the Haldane phase is not an intrinsically  topologically ordered phase,
but actually an example of SPT phase protected by translation and $SO(3)$ spin
rotation symmetries. This result is supported by a recent realization that the
existence of the Haldane phase requires symmetry (such as parity, time
reversal, or spin rotational symmetry).\cite{GuWen,pollmann} In other words, if
the necessary symmetries are absent, the Haldane phase can continuously connect
to the trivial phase without any phase transition.  We would like to mention
that, strictly speaking, the topological
insulators\cite{KM0501,BZ,KM0502,MB0706,FKM0703,QHZ0837} are not intrinsically
topologicaly ordered phases either. They are other examples of SPT phases
protected by time reversal symmetry.

Later, Haldane and
his collaborators discovered that the entanglement spectrum of the
Haldane phase (and other nontrivial phases) is doubly
degenerate.\cite{entanglspect} The entanglement spectrum degeneracy
is symmetry-protected.  A systematic classification of all SPT
orders for all 1D gapped systems was obtained in
Ref.~\onlinecite{CGW2} using projective representations.  All 1D
gapped phases are either symmetry breaking phases or SPT phases.

\begin{figure}[htbp]
\centering
\includegraphics[width=2.6in]{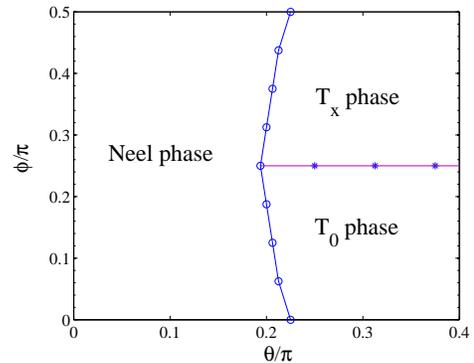}
\caption{(Color online) The phase diagram of model (\ref{HT0Tx}).
The transition between the Neel phase and the SPT phases are second
order, and the transition between $T_0$ and $T_x$ is first order. }
\label{fig:phase dgm-T0Tx}
\end{figure}

In this paper, we will study the SPT phases of a spin-chain protected by
translational symmetry and on-site $D_{2h}=D_2\otimes Z_2$ symmetry,
where the $Z_2$ in $D_2\otimes Z_2$ is generated by the time
reversal symmetry $T$.  Here we assume that the physical spin forms
a linear representation of $D_{2}$ and $T^2=1$.
First, we study a simple spin-1 model with those symmetries
\begin{eqnarray}
\label{HT0Tx}
H&=&\sum_i\Big[ \cos\theta S_{x,i}S_{x,i+1} +
\sin\theta[\cos\phi(S_{y,i} S_{y,i+1}
\\&&
+ S_{z,i} S_{z,i+1}) + \sin\phi(S_{xz,i}
S_{xz,i+1} + S_{xy,i}S_{xy,i+1})]\Big].
\nonumber
\end{eqnarray}
where $S_{mn}=S_m S_n+S_n S_m$ $(m,n=x,y,z)$. As shown in
Fig.~\ref{fig:phase dgm-T0Tx}, this model has three phases, the Neel
phase, the $T_0$ phase and the $T_x$ phase. The Neel phase breaks
the $D_{2h}$ symmetry, and the other two phases do not break any
symmetry. We can use sublattice spin magnetization as an order
parameter to distinguish the Neel phase. However, the remaining two
phases cannot be distinguished through local order parameter such as
sublattice spin magnetization. Further, in the both phases, the
entanglement spectrum\cite{entanglspect} is doubly degenerate, so
both of them are nontrivial. However, the entanglement spectrum is
not a good order parameter to separate them. Therefore, we need a
new tool to distinguish these two non-trivial phases that cannot be
described by symmetry breaking and the entanglement spectrum. It
turns out that the new tool is the projective representation of the
symmetry group. The two phases can be distinguished since their
doubly degenerate end states form different projective
representations of $D_{2h}$. The physical consequence is that they
response differently to weak external magnetic field.
The doubly degenerate end states can be viewed as an effective
spin-1/2 spin with asymmetric $g$-factors: $g_x$ and $g_y$ and $g_z$
describing the coupling of the end spin to external magnetic field
in $x$, $y$ and $z$ directions. We find that $g_x, g_y, g_z\neq 0$
in the $T_0$-phase and $g_x \neq 0, g_y=g_z = 0$ in the $T_x$-phase.
We would like to stress that such a property is robust against any
perturbations that do not break the $D_{2h}$ symmetry (the
perturbation may even break the translation symmetry).

The $D_{2h}$ symmetric spin-chain can have very rich quantum phases. It is
shown that it can have $64$ different gapped phases that do not break the
$D_{2h}$ and the translation symmetry.\cite{CGW3} In fact, it is the projective
representation theory that allow one to find all the non-trivial SPT phases
beyond the symmetry breaking description.  In this paper, we will not study all
of them. We will only use 8 classes of projective representations of the
$D_{2h}$ group to construct 8 gapped no-symmetry-breaking phases, one is
trivial and the other 7 are nontrivial SPT phases.  We find that four of the 7
SPT phases (labeled as $T_0, T_x, T_y$ and $T_z$) can be realized in $S=1$ spin
chain.  Here $T_0$ is the usual Haldane phase (because it includes the
Heisenberg point), and $T_x, T_y, T_z$ are new SPT phases.  The states in
different SPT phases cannot be smoothly connected to each other without
explicitly breaking the $D_{2h}$ symmetry in the Hamiltonian. The remaining
three SPT phases cannot be realized for $S=1$ chains, but they may exist in
$S=1$ spin ladders or $S=2$ spin chains.

The four SPT phases are experimentally distinguishable due to the
different behaviors of their end states.  In the $T_0$ phase, the
end states can be considered as spin-1/2 free spins. So weak
magnetic field couples to the end spins and lifts the ground state
degeneracy at linear order. However, in $T_x$ phase, the end states
can no longer be considered as normal spin-1/2 spins because they
behave differently under time reversal. As mentioned above, the
$g$-factors $g_y,g_z\approx0$, which
means that $B_y$ and $B_z$ can not split the degenerate ground
states in $T_x$ phase at linear order. Similarly, the end states of
$T_y$ (or $T_z$) only respond to $B_y$ (or $B_z$). According to
these properties, we propose an experimental scenario to distinguish
these four phases.

This paper is organized as the following. In section II, we
introduce the four SPT phases for the $S=1$ spin chain models. In
section III, we focus on the interaction of the end states to weak
external magnetic fields and propose an experimental method to
distinguish different SPT phases. In section IV we briefly summarize
the relationship between the SPT phases and the classes of
projective representations, and leave detailed derivations to the
appendix. Section V is devoted to conclusions and discussions.

\section{The model and SPT phases}

$D_{2h}$ group has eight group elements, $D_{2h}=\{E, R_x,R_y,R_z,
T, R_xT, R_yT, R_zT\}$, which is a direct product of the 180$^\circ$
spin rotation group $D_2=\{E, R_x=e^{-i\pi S_x},R_y=e^{-i\pi
S_y},R_z=e^{-i\pi S_z}\}$ and time reversal symmetry group
$Z_2=\{E,T\}$. Note that $T$ inverts the spin $(S_x,S_y,S_z)\to
(-S_x,-S_y,-S_z)$ and is anti-unitary. $D_{2h}$ has eight 1-d linear
representations (as shown in Tab.~\ref{tab:Repd2h} in appendix
\ref{appB}). Since $T$ is anti-unitary, the bases $|\phi\rangle$ and
$i|\phi\rangle$ have different time reversal parity. This subtle
property yields more than one SPT phases.

The most general Hamiltonian for an $S=1$ spin chain with $D_{2h}$
symmetry and with only nearest neighbor interaction is given by
\begin{eqnarray}\label{Hd2h}
H_{D_{2h}}&=&\sum_{i}[a_{1}S_{x,i}^2S_{x,j}^2+a_{2}(S_{x,i}^2S_{y,j}^2+S_{y,i}^2S_{x,j}^2)
\nonumber\\&&+a_{3}S_{y,i}^2S_{y,j}^2+a_{4}(S_{x,i}^2S_{z,j}^2+S_{z,i}^2S_{x,j}^2)
\nonumber\\&&+a_{5}S_{z,i}^2S_{z,j}^2+a_{6}(S_{y,i}^2S_{z,j}^2+S_{z,i}^2S_{y,j}^2)
\nonumber\\&&+b_{1}S_{x,i}S_{x,j}+b_{2}S_{yz,i}S_{yz,j}+
c_{1}S_{y,i}S_{y,j}\nonumber\\&&+c_{2}S_{xz,i}S_{xz,j}+
d_{1}S_{z,i}S_{z,j}+d_{2}S_{xy,i}S_{xy,j}\nonumber\\&&+e_1S_{x,i}^2+e_2S_{y,i}^2+e_3S_{z,i}^2].
\end{eqnarray}
where $j=i+1$, $S_{mn}=S_m S_n+S_n S_m$ $(m,n=x,y,z)$ and
$a_1,a_2,...,e_1,e_2,e_3$ are constants. We are interested in the
parameter regions within which the excitations are gapped and the
ground states respect the $D_{2h}$ symmetry.

In general, for 1D systems with translation symmetry and on-site
symmetry group $G$, a gapped ground state that does not break any
symmetry can be approximately written as a matrix product state
(MPS)
\begin{eqnarray}\label{MPS}
|\psi\rangle=\sum_{\{m_1,...,m_N\}}
\mathrm{Tr}(A^{m_1}...A^{m_N})|m_1...m_N\rangle,
\end{eqnarray}
which varies in the following way under the symmetry group
\begin{eqnarray}\label{MAM}
\sum_{m'}u(g)_{mm'}A^{m'}=\alpha(g)M(g)^\dag A^m M(g)
\end{eqnarray}
where $g\in G$ is a group element, and $\alpha(g)$/$M(g)$ is its
linear/projective representation matrix. Thus it is concluded that
the SPT phases are classified by $(\omega,\alpha)$, where $\omega$
is the element of the second cohomology group $H^2(G,\mathbb{C})$
(which describe different classes of projective representations of
the symmetry group $G$).\cite{CGW2}

So in our case, the ground state can be generally written in forms
of MPS shown in Eq.~(\ref{MPS}). The requirement $|\psi\rangle$
being invariant under $D_{2h}$ is equivalent to the condition in
Eq.~(\ref{MAM}). The main task of this paper is to try to find
different kinds of states that satisfy this condition. In this
paper, we only consider the case $\alpha(g)=1$. The full
classification (with a different approach from this paper) is given
in Ref.~\onlinecite{CGW3}.

Let us first consider the on-site terms in (\ref{Hd2h}). When
$|e_1|$, $|e_2|$ or  $|e_3|$ is large, the ground state of
$H_{D_{2h}}$ is simple. For instance, when $e_3\to-\infty$, the
ground state is a long-range ordered state which breaks the $D_{2h}$
symmetry; when $e_3\to\infty$, the ground state is a product state
$|\psi\rangle=\prod_i\otimes|0\rangle_i$, which is
trivial.\cite{e1e2e3} Since we are interested in the nontrivial SPT
phases, we will set $e_1=e_2=e_3=0$ in the following discussion.

\subsection{Exactly solvable models}

The Affleck Kennedy Lieb Tasaki (AKLT) model\cite{AKLT} is an
exactly solvable model with $SO(3)$ symmetry that falls in the
Haldane phase. The AKLT model contains all the physical properties
of the Haldane phase and all other states in this phase can smoothly
deform into the AKLT state. Since the ground state wave function of
this exactly solvable model is a simple matrix product (sMP) 
state\cite{iMPS} and is known in advance, so studying this model is
relatively easy and helps to understand the physics of the Haldane
phase.

In this section, we will introduce four classes of exactly solvable
models that have $D_{2h}$ symmetry. Analogous to the AKLT state, the
ground states of these exactly solvable models are nontrivial sMP
states satisfying Eq.~(\ref{MAM}). We will show that different
classes of sMP states can not be smoothly connected, which indicates
that each class corresponds to a phase.

The first example is a direct generalization of the AKLT model. The
ground state of AKLT state is represented by $A^x=\sigma_x,\ A^y=
\sigma_y,\ A^z=\sigma_z$, which has $SO(3)$ symmetry. When
generalized to $D_{2h}$ symmetry, we obtain
\begin{eqnarray}\label{T0}
A^x=a\sigma_x,\ \ A^y=b\sigma_y,\ \ A^z=c\sigma_z,
\end{eqnarray}
where $a,b,c$ are nonzero real numbers (the same below). When
$a=b=c=1$, the above state reduces to the AKLT state. For this
reason, we say that this model also belongs to the Haldane phase. We
label this phase as $T_0$. Similar to the AKLT model, the parent
Hamiltonian of above state is composed by projectors (for details
see appendix \ref{appA} and \ref{appB})
\begin{eqnarray}\label{Ht0}
H_0&=&\sum_i\left[({1\over4}+b^2c^2\gamma )S_{x,i}S_{x,j} +
({1\over4}+a^2c^2\gamma )S_{y,i}S_{y,j} \right. \nonumber\\
&&\left. + ({1\over4}+a^2b^2\gamma )S_{z,i}S_{z,j}+
({1\over4}-{b^2c^2\gamma })S_{yz,i}S_{yz,j} \right. \nonumber\\
&&\left. + ({1\over4}-{a^2c^2\gamma })S_{xz,i}S_{xz,j}+
({1\over4}-{a^2b^2\gamma
})S_{xy,i}S_{xy,j}\right] \nonumber\\
&&\left.\right.+h_0.
\end{eqnarray}
where $\gamma={1\over2(a^4+b^4+c^4)}$ and
\begin{eqnarray*}
h_0&=& -\sum_i[c^4{\gamma }(S_{x,i}^2S_{y,j}^2+
S_{y,i}^2S_{x,j}^2)+{b^4\gamma
}(S_{x,i}^2S_{z,j}^2\\&&+S_{z,i}^2S_{x,j}^2)+{a^4\gamma
}(S_{y,i}^2S_{z,j}^2+ S_{z,i}^2S_{y,j}^2)].
\end{eqnarray*}

At open boundary condition, the Hamiltonian (\ref{Ht0}) has exactly
four-fold degenerate ground states independent of the chain length.
The above exactly solvable model is frustration free, that is, the
expectation value of the Hamiltonian is minimized locally in the
ground states. The excitations are gapped and all correlation
functions of local operators are short ranged. Furthermore, if
$a,b,c$ are normalized $a^2+b^2+c^2=1$, then it is easily checked
that
\begin{eqnarray*}
\sum_mA^m(A^m)^\dag=I,\ \  \sum_m(A^m)^\dag \Lambda^2 A^m=\Lambda^2,
\end{eqnarray*}
here $\Lambda=I$, indicating that the entanglement spectrum of the
ground states is doubly degenerate. This informs that state
(\ref{T0}) is nontrivial. Actually, the models at the vicinity of
(\ref{Ht0}) (the phase $T_0$) have very similar properties unless
gap closing (second order phase transition) or level crossing (first
order phase transition) happens.

Now we consider another example of sMP state,
\begin{eqnarray}\label{Tx}
A^x=ia\sigma_x,\ \ A^y=b\sigma_y,\ \ A^z=c\sigma_z.
\end{eqnarray}
Above sMP state is also invariant under $D_{2h}$ group. As will be
shown later, it can not be continuously connected to Eq.~(\ref{T0})
without breaking the $D_{2h}$ symmetry. This means that it belongs
to another phase which we label as $T_x$ phase. The parent
Hamiltonian of (\ref{Tx}) is given by
\begin{eqnarray}\label{Htx}
H_x&=&\sum_i\left[({1\over4}+b^2c^2\gamma )S_{x,i}S_{x,j} +
({1\over4}-a^2c^2\gamma )S_{y,i}S_{y,j} \right. \nonumber\\
&&\left. + ({1\over4}-a^2b^2\gamma )S_{z,i}S_{z,j}+
({1\over4}-{b^2c^2\gamma })S_{yz,i}S_{yz,j} \right. \nonumber\\
&&\left. + ({1\over4}+{a^2c^2\gamma })S_{xz,i}S_{xz,j}+
({1\over4}+{a^2b^2\gamma
})S_{xy,i}S_{xy,j}\right] \nonumber\\
&&\left.\right.+h_0.
\end{eqnarray}

Similarly, the third example
\begin{eqnarray}\label{Ty}
A^x=a\sigma_x,\ \ A^y=ib\sigma_y,\ \ A^z=c\sigma_z
\end{eqnarray}
belongs to the $T_y$ phase and its parent Hamiltonian is
\begin{eqnarray}\label{Hty}
H_y&=&\sum_i\left[({1\over4}-b^2c^2\gamma )S_{x,i}S_{x,j} +
({1\over4}+a^2c^2\gamma )S_{y,i}S_{y,j} \right. \nonumber\\
&&\left. + ({1\over4}-a^2b^2\gamma )S_{z,i}S_{z,j}+
({1\over4}+{b^2c^2\gamma })S_{yz,i}S_{yz,j} \right. \nonumber\\
&&\left. + ({1\over4}-{a^2c^2\gamma })S_{xz,i}S_{xz,j}+
({1\over4}+{a^2b^2\gamma
})S_{xy,i}S_{xy,j}\right] \nonumber\\
&&\left.\right.+h_0.
\end{eqnarray}

The last example
\begin{eqnarray}\label{Tz}
A^x=a\sigma_x,\ \ A^y=b\sigma_y,\ \ A^z=ic\sigma_z
\end{eqnarray}
belongs to the $T_z$ phase with its parent Hamiltonian given by
\begin{eqnarray}\label{Htz}
H_z&=&\sum_i\left[({1\over4}-b^2c^2\gamma )S_{x,i}S_{x,j} +
({1\over4}-a^2c^2\gamma )S_{y,i}S_{y,j} \right. \nonumber\\
&&\left. + ({1\over4}+a^2b^2\gamma )S_{z,i}S_{z,j}+
({1\over4}+{b^2c^2\gamma })S_{yz,i}S_{yz,j} \right. \nonumber\\
&&\left. + ({1\over4}+{a^2c^2\gamma })S_{xz,i}S_{xz,j}+
({1\over4}-{a^2b^2\gamma
})S_{xy,i}S_{xy,j}\right] \nonumber\\
&&\left.\right.+h_0.
\end{eqnarray}

Above we have given four special models that belong
to different SPT phases. In the next subsection, we will show that if
one keeps the $D_{2h}$ symmetry, phase transitions must happen when
connecting these models.

\subsection{transitions between different SPT phases}

In order to justify the four SPT phase, we will use numerical method
to study more general Hamiltonians. The method we adopt is one
version of the tensor renormalization group approach developed in 1D
by G. Vidal\cite{Vidal} and later generalized to 2D by T. Xiang
\textit{et.al}.\cite{TensRG} In this method, the ground state is
approximated by a MPS. For an arbitrarily initialized state, we can
act the (infinitesimal) imaginary time evolution operator
$U(\delta\tau)= e^{-H\delta\tau}$ for infinite times, finally
obtaining the fixed point matrix $A^m$. If the dimension $D$ of
$A^m$ is not too small, the corresponding MPS is very close to the
true ground state. In our numerical calculation, we set $D=16$. In
1D, the ground state energy, correlation functions, density matrix,
and entanglement spectrum can be calculated directly from the matrix
$A^m$.

\begin{figure}
  \centering
  \subfigure[]{
    \label{fig:E_smooth} 
    \includegraphics[width=1.6in]{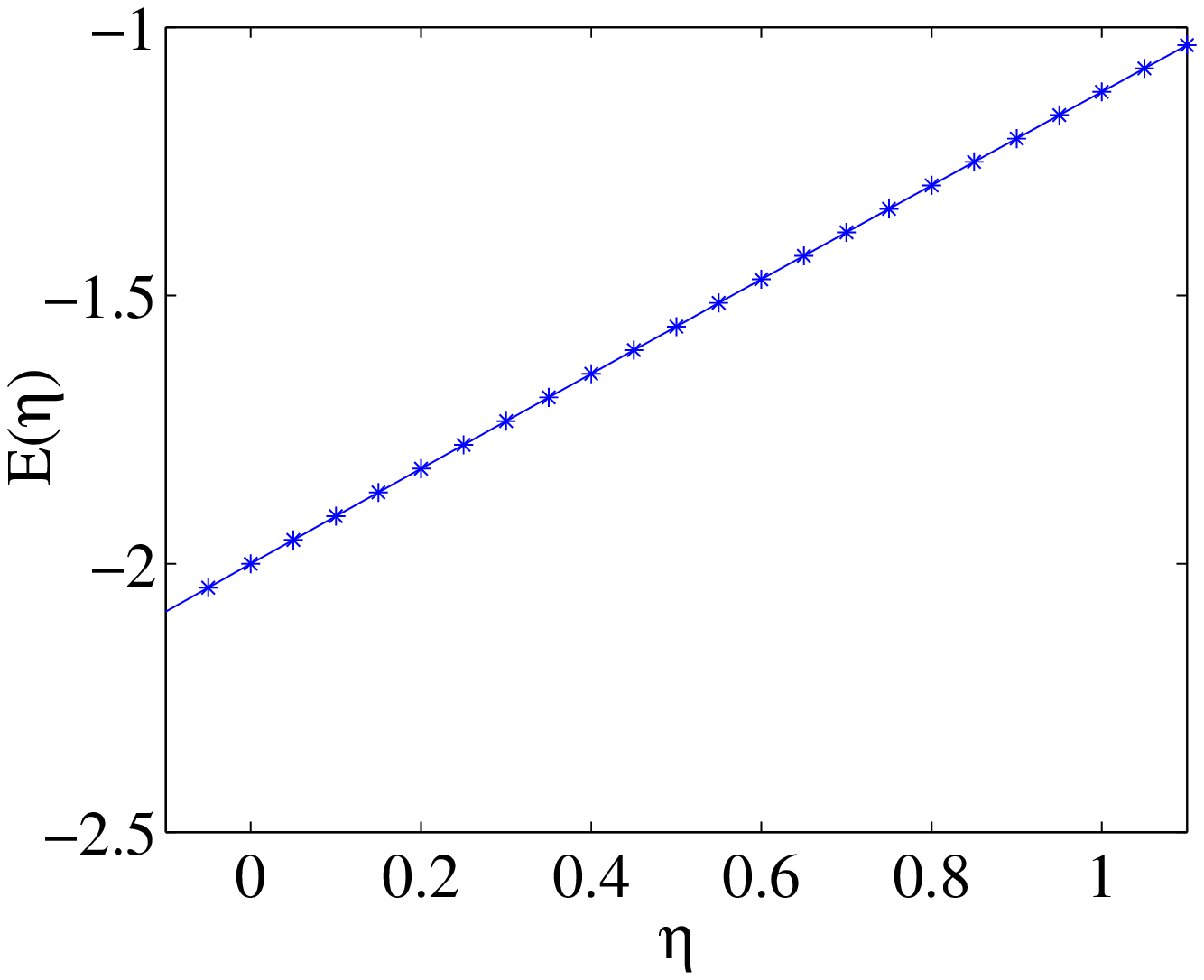}}
  \subfigure[]{
    \label{fig:ED_EDD_smooth} 
    \includegraphics[width=1.6in]{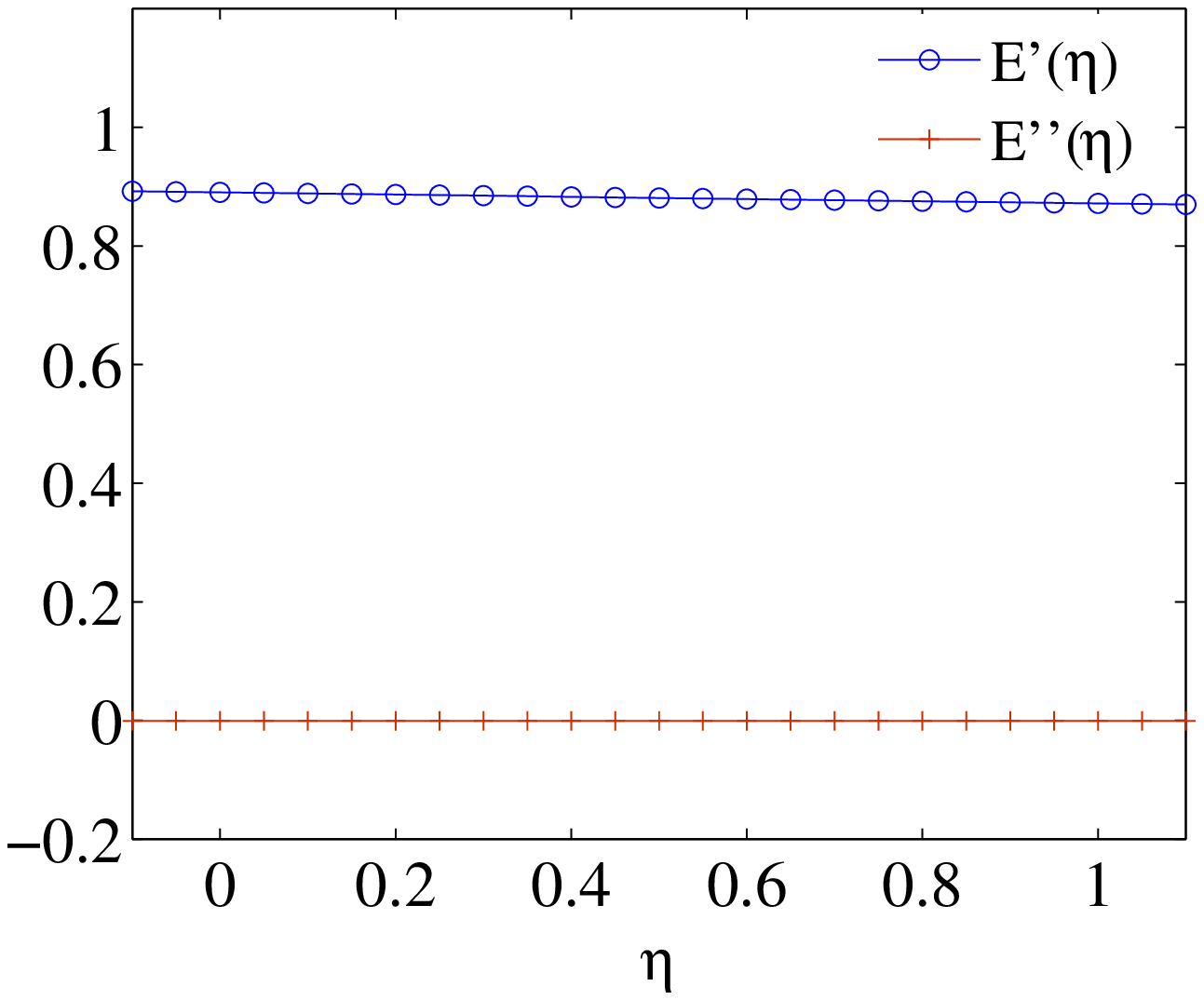}}
  \caption{(Color online) (a) The energy curve $E(\eta)$, (b) The first and second derivatives of
$E(\eta)$. All these curves are smooth, indicating that these states
are in the same phase.}
  \label{fig:smooth} 
\end{figure}

Noticing that $h_0$ is a common term in the four exactly solvable
models, which indicates that it is unimportant and can be dropped.
This can be numerically verified. For this purpose, we add a
perturbation to the models, such as (\ref{Htx}),
\begin{eqnarray}\label{eta}
H(\eta)=H_x-\eta h_0,
\end{eqnarray}
where $\eta\in[0,1]$. As shown in Fig.~\ref{fig:smooth}, the ground
state energy $E(\eta)$ and its derivatives $E'(\eta), E''(\eta)$ are
all smooth functions, indicating that all the Hamiltonians $H(\eta)$
belong to the same phase. Using the same method, one can also check
that the Hamiltonians (with $D_{2h}$ symmetry) in the vicinity of an
exactly solvable model fall in the same phase. For instance, the
Heisenberg model and $H_0$ in (\ref{Ht0}) are in the same phase.

Now a question is whether the ground states of different exactly
solvable models can be smoothly transformed into each other. For
this end, we consider a more realistic model
(\ref{HT0Tx})
which connects two
exactly solvable models, such as $H_0$ and $H_x$.
We are interested in the anti-ferromagnetic cases and will focus on
the parameter region $\theta,\phi\in[0,{\pi\over2}]$. The point
$({\pi\over4},0)$ is the Heisenberg model. From the result of the last
paragraph, the Heisenberg model is in the same phase as
(\ref{Ht0}), and similarly $({\pi\over4},{\pi\over2})$ is in the
same phase as (\ref{Htx}). If these two points cannot be smoothly
connected (i.e.,if gap closing or level crossing will unavoidably
happen), then (\ref{Ht0}) and (\ref{Htx}) belong to different
phases.

\begin{figure}
  \centering
  \subfigure[]{
    \label{fig:E_phi=1ov8} 
    \includegraphics[width=2.4in]{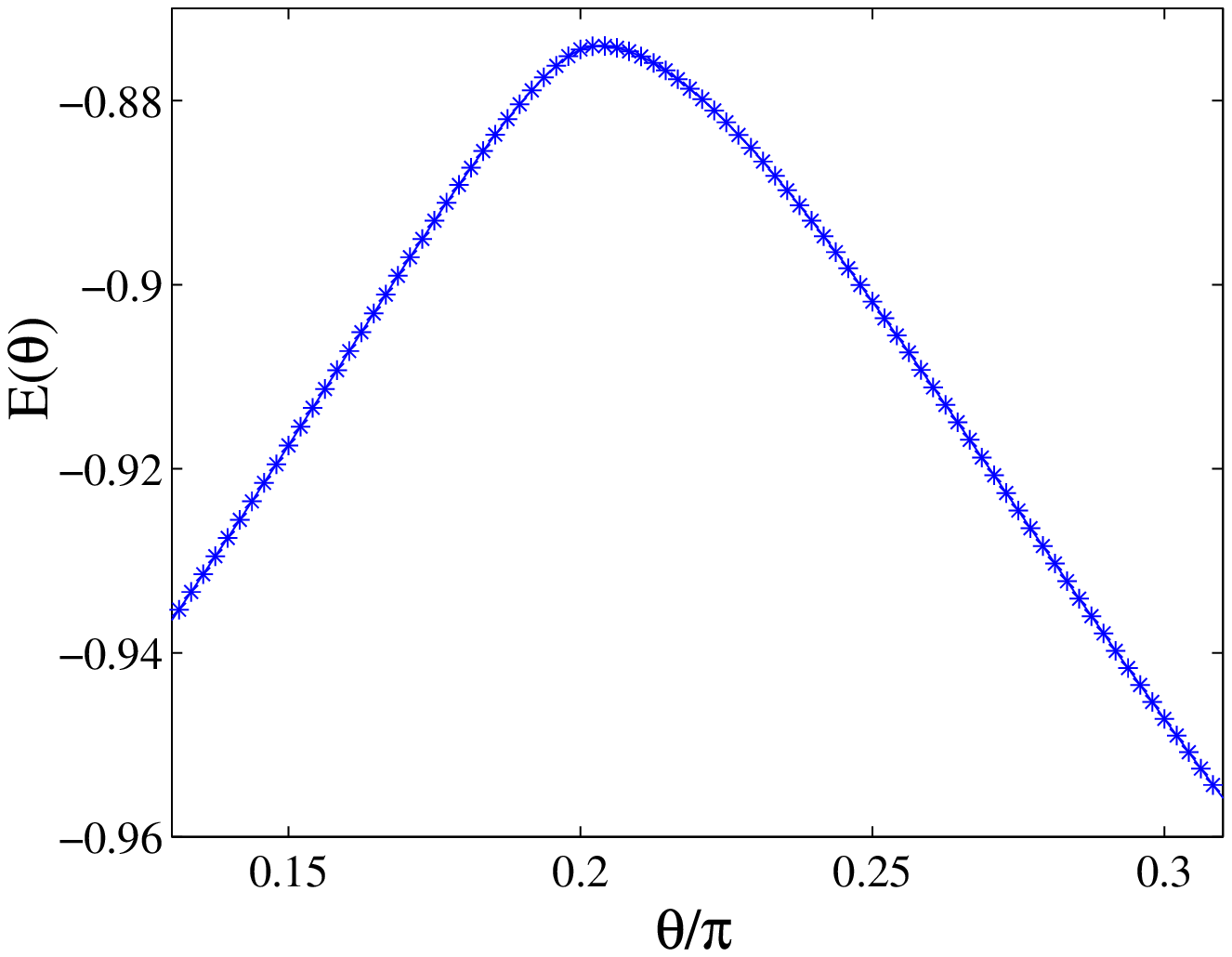}}
  \subfigure[]{
    \label{fig:ED_EDD_phi=1ov8} 
    \includegraphics[width=2.4in]{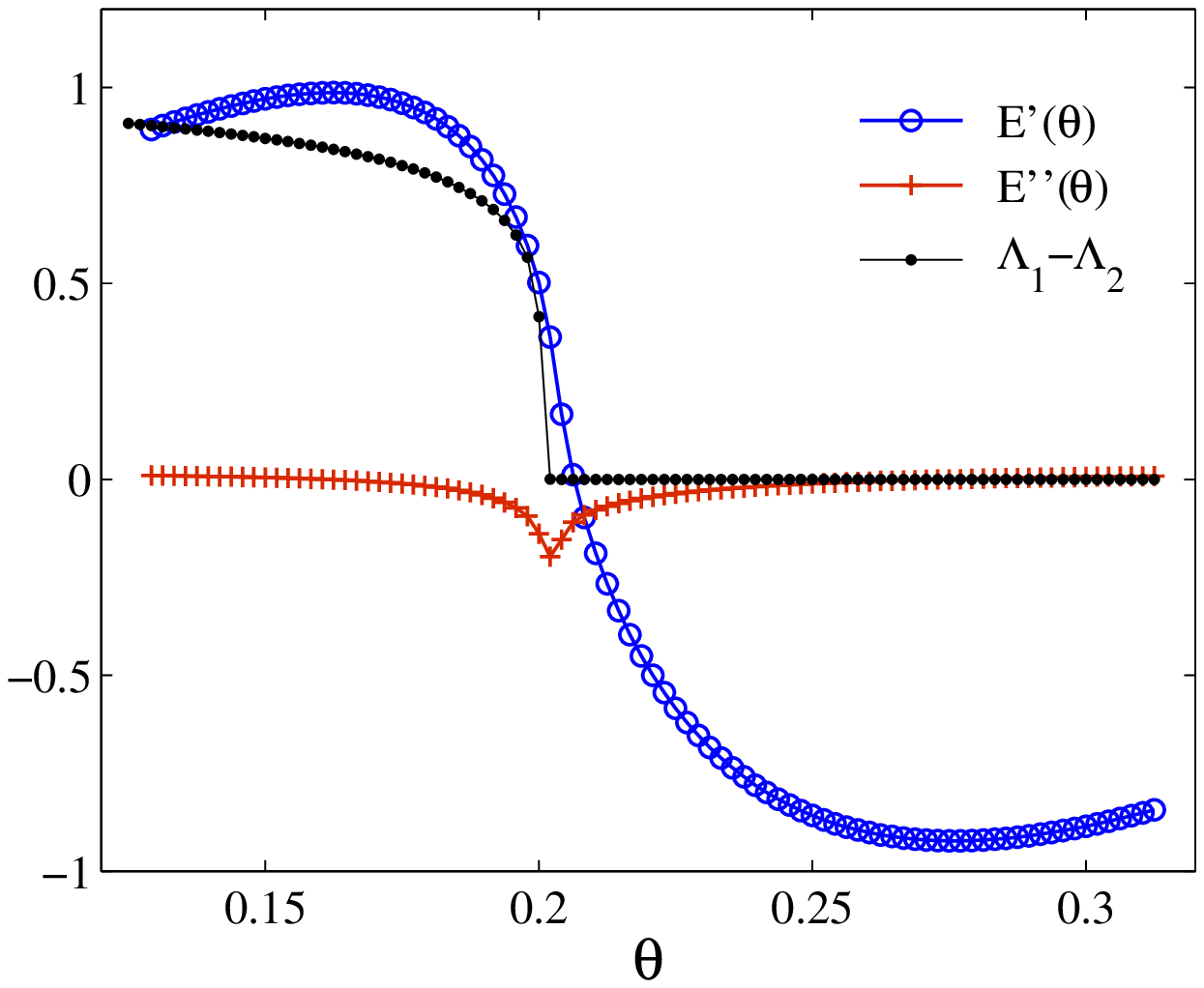}}
  \caption{(Color online) (a) The energy curve $E(\theta)$ with fixed
$\phi={\pi\over8}$ (and $\phi={3\pi\over8}$ ). (b) The line with
circles is the first order derivative and the line with crosses is
the second order derivative of $E(\theta)$; the doted line is the
difference tween the two biggest weights of the entanglement
spectrum $\Lambda_1-\Lambda_2$. Since $E'(\theta)$ is continuous but
$E''(\theta)$ is not, the transition is second order. The degeneracy
of $\Lambda_1$ and $\Lambda_2$ indicates that the $T_0$ (and $T_x$)
phase is nontrivial. }
  \label{fig:phi=1ov8} 
\end{figure}

Using the tensor RG method, we can calculate the ground state energy
of (\ref{HT0Tx}) and the phase diagram is shown in
Fig.~\ref{fig:phase dgm-T0Tx}. When $\theta$ is less then
${0.21\pi}$, the ground state is Neel ordered. When $\theta$
increases, a second order phase transition occurs and we enter the
SPT phases. Fig.~\ref{fig:phi=1ov8} shows the data of energy curve
with fixed $\phi={\pi\over8}$ (and $\phi={3\pi\over8}$) and its
first and second derivatives , which illustrate this transition.

%

The region $\phi\in[0,{\pi\over4})$ belongs to the $T_0$ phase and
$\phi\in({\pi\over4},{\pi\over2}]$ belongs to the $T_x$ phase. A
first order phase transition between them happens at
$\phi={\pi\over4}$.

The first order phase transition occurs exactly at
$\phi={\pi\over4}$. This is because the whole phase diagram is
symmetric above and below the line $\phi={\pi\over4}$. This symmetry
can be seen in the Hamiltonian. Notice that under a unitary matrix
$U$, we get
\begin{eqnarray}\label{SU(2)}
&U^\dag S_xU= S_{x},\ \ U^\dag S_yU=-S_{xz},\ \ U^\dag S_zU=
S_{xy},\end{eqnarray}
where $U=\left(\begin{matrix}{1\over\sqrt2}e^{i{\pi\over4}}&0&{1\over\sqrt2}e^{-i{\pi\over4}}\\0&1&0\\
{1\over\sqrt2}e^{-i{\pi\over4}}&0&{1\over\sqrt2}e^{i{\pi\over4}}\end{matrix}\right)$.
\cite{SU(2)} This means that Hamiltonian (\ref{HT0Tx}) satisfies
$(\prod_i\otimes U_i)^\dag H(\theta,\phi)(\prod_i\otimes U_i)=
H(\theta, {\pi\over2}-\phi)$, which yields $E(\theta,\phi) =
E(\theta,{\pi\over2}-\phi)$, and their ground states are transformed
by the (local) unitary transformation $\prod_i\otimes U_i$. But this
unitary transformation is not invariant under time reversal $T$, so
the behavior of the ground state also changes under time reversal.
Resultantly, the state after the transformation belongs to a
different phase.

The entanglement spectrum of the ground states can also be obtained
programmatically. We find that in both $T_0$ and $T_x$ phases the
entanglement spectrum is doubly degenerate (see
Fig.~\ref{fig:ED_EDD_phi=1ov8}). This shows that the $T_0$ and $T_x$
phases are indeed nontrivial. Similar to the model (\ref{HT0Tx}),
the phase transition between $T_0$ and $T_y$ or between $T_0$ and
$T_z$ can also be illustrated.

\begin{figure}[htbp]
\centering
\includegraphics[width=2.6in]{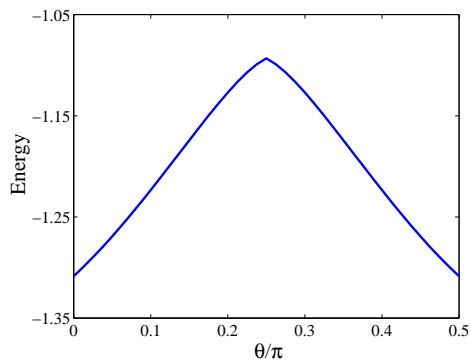}
\caption{(Color online) The ground state energy for model
(\ref{HTxTz}). A first order phase transition is obvious.}
\label{fig:TxTz}
\end{figure}

Now we will show that first order phase transition also exists
between any two of $T_x, T_y, T_z$. As an example, we consider the
model that contains the transition between $T_x$ and $T_z$ phases:
\begin{eqnarray}\label{HTxTz}
H&=&\sum_i[{1\over6}S_{y,i}S_{y,j} + {5\over6}S_{xz,i} S_{xz,j} +
\cos\theta (S_{x,i}S_{x,j}\nonumber\\&& + S_{xy,i}S_{xy,j})
+\sin\theta(S_{z,i} S_{z,j} + S_{yz,i} S_{yz,j})].\nonumber\\
\end{eqnarray}
When $\theta=\tan^{-1}{1\over5}$, the above Hamiltonian is in the
same phase as $H_x$ as shown in (\ref{eta}), and when
$\theta=\tan^{-1}5$, it is in the same phase as $H_z$. The ground
state energy of (\ref{HTxTz}) as a function of $\theta$ can be
obtained using the tensor RG method, and the result is shown in
Fig.~\ref{fig:TxTz}. A first order transition at
$\theta={\pi\over4}$ manifests itself. For the reason similar to
(\ref{SU(2)}), the model also has a symmetry
$E(\theta)=E({\pi\over2}-\theta)$.

From the above analysis, we can conclude that the four exactly solvable
models really stand for four distinct SPT phases. All these
SPT phases are protected by the $D_{2h}$ symmetry. As will
be shown in section IV, no more SPT phases exist for $S=1$
models with $D_{2h}$ symmetry. Furthermore, the Eqs.~(\ref{HT0Tx})
and (\ref{HTxTz}) show that these SPT phases can be obtained
by much simpler Hamiltonians which is hopefully realized
experimentally.

Now an interesting question arises, how to distinguish these SPT
phases in a practical way? It is impossible to distinguish these
phases by linear response in the bulk since it is gapped. However,
the end `spins' localized at open boundaries may have different
behaviors in different SPT phases. In the next section, we will
propose an experimental method to detect each SPT phase.

\section{Distinguishing different SPT phases}
We expect to distinguish the four SPT phases through their different
physical properties. Experimentally, all measurable physical
quantities are response functions, or susceptibilities. So we need
to add small perturbations and expect that (the `end spins' of)
different phases have different responses. The simplest perturbation
for spin system is magnetic field $H'=g_L\mu_B\mathbf B\cdot\tilde
{\mathbf S}$, here $\tilde {\mathbf S}=\sum_i\mathbf S_i$, $g_L$ is
the Lande factor and $\mu_B$ is the Bohr magneton. We will study the
linear response to small $\mathbf B$.

Since the states in the same phase have the same universal
properties, we will focus on the exactly solvable models first. For
simplicity, we consider the AKLT model, namely, $H_0$ with
$a=b=c=1$. In the matrix product state picture, the physical $S=1$
spin is divided into two $J=1/2$ virtual spins. In the AKLT state,
the virtual spins pair into singlets (called valence bonds) on each
link between neighboring sites. Under open boundary condition, a
free $J=1/2$ spin at each end remains unpaired. The two end spins
account for the exactly four-fold degeneracy of the ground states.
In this picture, it's easy to calculate the total spin in the ground
state Hilbert space. The singlets in the bulk have no contributions
to $\tilde {\mathbf S}$, only the two end spins $\mathbf j_1$ and
$\mathbf j_2$ contribute and resultantly $\tilde {\mathbf S}=\mathbf
j_1+\mathbf j_2$. In this sense, the end spins can be considered as
\textit{impurity spins} of a paramagnetic material. Since the total
spin of an open chain is ${1\over2}\otimes{1\over2}=0\oplus1$, we
expect that the eigenvalue of $\tilde S_x, \tilde S_y, \tilde S_z$
should be $1,-1,0,0$. This can be verified by exact diagonalizing a
short chain. We denote these four degenerate ground states as
$|\psi_1\rangle, |\psi_2\rangle, |\psi_3\rangle, |\psi_4\rangle$.
Then the matrix element of $\tilde S_m$ in the ground state Hilbert
space is given by
\begin{eqnarray}
\tilde S_m^{\alpha\beta}=\langle\psi_\alpha|\tilde
S_m|\psi_\beta\rangle,\ \ \alpha,\beta=1,2,3,4¡£
\end{eqnarray}
The eigenvalues of the matrices $(\tilde {\mathbf S}^{\alpha\beta})$
are exactly $1,-1,0,0$ and these values are independent of the
length of the chain. Thus a small magnetic field along any direction
$H'=g_L\mu_B B_xS_x$ or $g_L\mu_B B_yS_y$ or $g_L\mu_B B_zS_z$ will
split the ground state degeneracy and give rise to a finite
magnetization.

At finite temperature, the susceptibility satisfies the Curie law
and is given by\cite{Kittle}
\begin{equation}
\label{Claw} \chi(T)\simeq {Ng^2\mu_B ^2\over3k_BT} ,
\end{equation}
where $N$ is the number of end spins, $g=\sqrt{J(J+1)}g_L$, $J=1/2$
and $g_L$ is the Lande $g$ factor. If the spin-1 chains in the
sample are broken into long separate segments, then $N$ can be a
considerable number. We also note that, in real samples, the
susceptibility also contains a temperature independent part coming
from the bulk.

We see that, for the AKLT model, the spin susceptibility diverges at
low temperature along all directions. For a general model in the
$T_0$ phase, the divergence of $\chi_x(T),\chi_y(T),\chi_z(T)$ still
holds, except that it is no longer isotropic.

\begin{figure}[htbp]
\centering
\includegraphics[width=3.in]{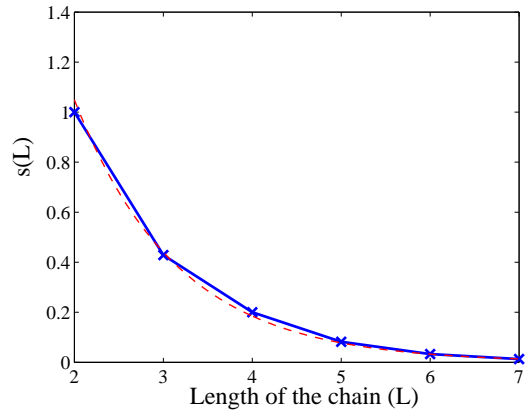}
\caption{(Color online) The eigenvalues of $\tilde S_y$ and $\tilde
S_z$ are $s,-s,0,0$. The magnitude of $s$ exponentially decays with
the length of the chain in $T_x$ phase. The dashed line is an
exponential fit. The results are obtained by exact diagonalization
and we only calculate up to seven sites. } \label{fig:SyinTx}
\end{figure}

However, in phase $T_x$, the end `spins' have absolutely different
physical properties. We consider the model $H_x$ in (\ref{Htx}), and
set $a=b=c=1$. Then we calculate the eigenvalues of operators
$\tilde S_x, \tilde S_y$ and $\tilde S_z$ in the ground state
Hilbert space as before. We find that the eigenvalues of $(\tilde
S^{\alpha\beta})_x$ are still $1,-1,0,0$, meaning that along $x$
direction the spin-1/2 end spins still exist and $\chi_x(T)$
diverges at $T=0$. The eigenvalues of $(\tilde S^{\alpha\beta})_y$
and $(\tilde S^{\alpha\beta})_z$ also have the structure $s,-s,0,0$,
but the magnitude of the nonzero eigenvalues $s$ exponentially decay
to zero with the increasing of the length of the chain (see
Fig.~\ref{fig:SyinTx}). This means that in $y$- and $z$-directions,
there are no free spins coupled to the magnetic field. In appendix
\ref{appD} we will show that this property is determined by the
projective representation carried by the virtual `spins'. In this
case, $\chi_y(T)$ and $\chi_z(T)$ are given by (\ref{Claw}) with
$g_y,g_z \approx 0$. The result that $\chi_x(T)$ follows Curie law
and $\chi_y(T)$, $\chi_z(T)$ has effective $g_y, g_z\approx 0$ is a
universal property of all the models in the $T_x$ phase.

Similarly, one can check that only $\chi_y(T)$ follows Curie law in
$T_y$ phase with the usual $g_y\approx\sqrt{J(J+1)}g_L$ ($g_x,g_z
\approx 0$), and similarly only $\chi_z(T)$ follows Curie law in
$T_z$ phase with the usual $g_z\approx\sqrt{J(J+1)} g_L$ ($g_x,g_y
\approx 0$). Therefore, by measuring the temperature dependence of
susceptibility and the effective $g$ in $x, y, z$ directions, we are
able to distinguish the four SPT phases.

\section{projective representations and SPT phases}

In previous sections, we have given four SPT phases of the
model (\ref{Hd2h}) and studied their physical properties. In this
section, we will explain how the $D_{2h}$ symmetry supports the
existence of these phases. Then we will discuss other possible
SPT phases of spin systems with $D_{2h}$ symmetry.

The ground state of a gapped phase is written as (\ref{MPS}). If we
require that the ground state MPS be invariant under the symmetry
group $D_{2h}$, namely, $g|\psi\rangle=|\psi\rangle$ ($g\in
D_{2h}$), then under the action of the symmetry group the matrix
$A^m$ must vary in the following way.

1) $g$ is unitary, $g\in\{E, R_x, R_y,R_z\}$,
\begin{eqnarray}\label{unitary}
\sum_{m'}u(g)_{mm'}A^{m'}=M(g)^\dag A^m M(g);
\end{eqnarray}

2) $g$ is anti-unitary, $g\in\{T, R_xT, R_yT,R_zT\}$,
\begin{eqnarray}\label{anti-unitary}
\sum_{m'}u(g)_{mm'}KA^{m'}=M(g)^\dag A^m M(g).
\end{eqnarray}
Here $u(g)$ and $M(g)$ are representations of the symmetry group
$D_{2h}$. The matrices $u(g)$ satisfy the same multiplication law of
$D_{2h}$ group and is called a \textit{linear representation}. The
physical spin freedoms are linear representations of $D_{2h}$.
$M(g)$ and $M(g)e^{i\theta}$ are equivalent and belong to the same
presentation. Up to a phase factor (which depends on the group
elements), $M(g)$ satisfy the multiplication law of $D_{2h}$, and
this kind of presentation is called a \textit{projective
representation}. The virtual `spins' (and the end states) are
projective representations of $D_{2h}$. More knowledge about
projective representation can be found in Ref.~\onlinecite{CGW2,
Boyal}.

The $D_{2h}$ group has eight 1-d linear representations (see
Tab.~\ref{tab:Repd2h}) and eight (and only eight) classes of
projective representations labeled by (111),
(11-1),(1-11),(1-1-1),(-111),(-11-1),(-1-11) and (-1-1-1) (see
Tab.~\ref{tab:prjRep}). The first class of projective representation
(111) is the eight 1-d linear representations, which are trivial (an
example of the the corresponding trivial phase is the case
$e_3\to\infty$). The other seven projective representations are 2-d
and nontrivial. Since states supporting different projective
representations (or virtual `spins') cannot be smoothly transformed
into each other, each projective representation corresponds to a SPT
phase. This means that there should be at least seven different
nontrivial SPT phases for spin systems respecting $D_{2h}$ symmetry.
(In fact, when considering different $\alpha(g)$ there are more
nontrivial SPT phases for spin systems respecting $D_{2h}$
symmetry.\cite{CGW3})

How can we obtain the projective representations? Mathematically,
finding the projective representations of a group $G$ is equivalent
to find the linear representations of its cover group, which is a
central extension of $G$ and is called representation group $R(G)$.
\cite{Boyal} The representation group $R(D_{2h})$ is available in
literature, \cite{Boyal} so we can calculate the matrix elements of
all the projective representations of $D_{2h}$ (see
Tab.~\ref{tab:prjRep}).

Once the matrices of the projective representations are obtained, we
can calculate the CG coefficients for decomposing the direct product
of two projective representations. From the CG coefficients, we can
construct sMP states and their parent Hamiltonians.\cite{Tu} The
models (\ref{Ht0}),(\ref{Htx}),(\ref{Hty}) and (\ref{Htz}) are
constructed accordingly and correspond to the
(-1-1-1),(-1-11),(-11-1),(-111) representations respectively. From
these models we can know what kind of interactions are essential for
each SPT phase.

As shown in appendix \ref{appB}, the remaining three SPT phases of
(1-11),(11-1),(1-1-1) cannot be realized for $S=1$ spin chains. The
reason is that the physical freedom is not sufficient to support the
direct product of two such projective representations. However,
these phases might be realized in $S=1$ spin ladders or $S=2$
models, and this will be our upcoming work.

\section{conclusion and discussion}

In summary, we have found four nontrivial SPT phases $T_0, T_x, T_y,
T_z$ of $S=1$ spin chains which have on-site $D_{2h}$ symmetry.
These SPT phases have similar properties as the usual Haldane phase,
such as the bulk excitation gap, short-range correlations, existence
of end `spins', entanglement spectrum degeneracy.  However, the
different projective representations of the end spin under $D_{2h}$
indicate that they do belong to different phases. The SPT order that
distinguishes them is the class of projective representations (or
the group elements of the second cohomology $H^2(D_{2h},\mathbb C)$)
correspond to the ground states (or the matrices $A^m$).

We find that different SPT phases can be distinguished by
experimental method. The magnetic susceptibilities $\chi_x, \chi_y,
\chi_z$ obey Curie law and diverge at zero temperature.  In $T_0$
phase the effective $g$-factors of the end spin have the usual
values for magnetic field in $x$-, $y$-, and $z$-directions.  But in
$T_x$ (or $T_y$ or $T_z$) phase, the effective
$g\approx\sqrt{J(J+1)}g_L$ has the usual value only for magnetic
field in $x$-direction (or $y$-direction or $z$-direction).  The
effective $g\approx 0$ [see eq. (\ref{Claw})] for magnetic field in
the other two directions.

From the seven nontrivial projective representations for $D_{2h}$
group, we constructed seven SPT phases.  Four of them are discussed
above. The other three may be realized in $S=1$ spin ladders or spin
$S=2$ models and are not discussed in the current paper.  Some
conclusion in this paper can be generalized to larger spin systems
and higher dimensions.

\section{acknowledgements}
We thank Xie Chen, Hong-Hao Tu, Zheng-Yu Weng and Hai-Qing Lin for helpful
discussions. This research is supported by  NSF Grant No.
DMR-1005541 and NSFC 11074140.

\appendix
\section{Spin chian with $D_2$ symmetry}\label{appA}

In appendix A, we will first study $S=1$ spin systems with $D_2$
symmetry. The same method can be applied to $D_{2h}$ case.

\subsection{General Hamiltonian with $D_2$ point group symmetry}

The point group $D_2$ has only four elements, $D_2=\{E, R_x, R_y,
R_z\}$. The multiplication table is shown in Tab.~\ref{tab:Mul}.
\begin{table}[htbp]
\caption{Multiplication table of $D_2$} \label{tab:Mul}
\begin{ruledtabular}
\begin{tabular}{c|cccc}
         & $E$        & $R_x$        & $R_y$      & $R_z$\\
\hline
$E$      & $E$        &      $R_x$   &   $R_y$    &    $R_z$    \\
$R_x$    & $R_x$      &       $E$    &   $R_z$    &    $R_y$    \\
$R_y$    & $R_y$      &      $R_z$   &   $E$      &    $R_x$    \\
$R_z$    & $R_z$      &      $R_y$   &  $R_x$     &    $E$
\end{tabular}
\end{ruledtabular}
\end{table}

It has four 1-d linear representations, whose matrix elements and
bases of representations are shown in Tab.~\ref{tab:Rep}. From
quantum mechanics, we know that the $2S+1$ bases of integer spin-$S$
span an irreducible linear representation space of $SO(3)$ group.
This Hilbert space is reduced into a direct sum of $2S+1$ 1-d
irreducible linear representation spaces of $D_2$. For example, when
$S=1$ (a vector), the bases
\begin{eqnarray*}
&&|x\rangle={1\over\sqrt2}(|-1\rangle-|1\rangle),\nonumber\\
&&|y\rangle={1\over\sqrt2}i(|-1\rangle+|1\rangle),\nonumber\\
&&|z\rangle=|0\rangle
\end{eqnarray*}
form the $B_3,B_2,B_1$ representations of $D_2$ respectively.

\begin{table}[htbp]
\caption{Linear representations of $D_2$} \label{tab:Rep}
\begin{ruledtabular}
\begin{tabular}{c|cccc|ccc}
         & $E$        & $R_x$        & $R_y$      & $R_z$&&bases or operators&\\
\hline
$A$      &      1     &      1       &      1     &     1    &$|0,0\rangle$&&$S_x^2,S_y^2,S_z^2$\\
$B_1$    &      1     &       -1     &      -1    &     1    &$|1,z\rangle$&$S_z$&$S_{xy}$\\
$B_2$    &      1     &      -1      &      1     &    -1    &$|1,y\rangle$&$S_y$&$S_{xz}$\\
$B_3$    &      1     &      1       &      -1    &    -1    &$|1,x\rangle$&$S_x$&$S_{yz}$
\end{tabular}
\end{ruledtabular}
\end{table}
Here we focus on the $S=1$ model with nearest neighbor interaction.
The general Hamiltonian with $D_2$ symmetry is given by
\begin{eqnarray}\label{H_d2}
H_{D_2}&=&H_{D_{2h}}+f_{1}(S_{x,i}S_{yz,j}+S_{yz,i}S_{x,j})\nonumber\\&&+
f_{2}(S_{y,i}S_{xz,j}+S_{xz,i}S_{y,j})\nonumber\\&&+
f_{3}(S_{z,i}S_{xy,j}+S_{xy,i}S_{z,j}).
\end{eqnarray}
where $f_1,f_2,f_3$ are constants and $H_{D_{2h}}$ is given in
(\ref{Hd2h}). The above Hamiltonian  $H_{D_2}$ also has
translational symmetry and spacial inversion symmetry. The
additional $f_1,f_2,f_3$ terms are odd under time reversal and break
the $T$ symmetry of $H_{D_2}$. To study the SPT phases, we need to
obtain the projective representations of $D_2$.

\subsection{Projective representation  and CG coefficients of $D_2$}
From Ref.~\onlinecite{Boyal}, determining the projective
representation of a point group $G$ is equivalent to determining the
linear representation of its representation group(s) $R(G)$ (which
cover $G$ integer times). There are two non-isomorphism
representation groups of $D_2$, namely, $R_1(D_2)$ and $R_2(D_2)$,
both of which have two generators $P$, $Q$ and eight group elements.
Their multiplication tables are listed in Tabs.~\ref{tab:Rl} and
\ref{tab:R2}. In the following, we will mainly discuss the covering
group $R_1(D_2)$, and leave the discussion about $R_2(D_2)$ to the
end of this section.
\begin{table}[htbp]
\caption{Multiplication table of $R_1(D_2$). Notice that
$P^4=Q^4=E$, $P^2=Q^2$ and $QP=P^3Q$.} \label{tab:Rl}
\begin{ruledtabular}
\begin{tabular}{c|cccccccc}
         & $E$        &      $P^2$   & $P^3$      &    $P$   &    $Q$  &    $P^2Q$  &   $PQ$    &   $P^3Q$\\
\hline
$E$      & $E$        &      $P^2$   &   $P^3$    &    $P$   &    $Q$  &    $P^2Q$  &   $PQ$    &   $P^3Q$\\
$P^2$    & $P^2$      &      $E$     &   $P$      &    $P^3$ &   $P^2Q$&    $Q$     &  $P^3Q$   &   $PQ$ \\
$P^3$    & $P^3$      &      $P$     &   $P^2$    &    $E$   &  $P^3Q$ &   $PQ$     &  $Q$      &   $P^2Q$\\
$P$      & $P$        &      $P^3$   &    $E$     &    $P^2$ &   $PQ$  &    $P^3Q$  &  $P^2Q$   &   $Q$  \\
$Q$      & $Q$        &      $P^2Q$  &    $PQ$    &    $P^3Q$&   $P^2$ &   $E$      &  $P$      &   $P^3$  \\
$P^2Q$   & $P^2Q$     &      $Q$     &   $P^3Q$   &    $PQ$  &   $E$   &   $P^2$    &  $P^3$    &   $P$ \\
$PQ$     & $PQ$       &      $P^3Q$  &   $P^2Q$   &    $Q$   &   $P^3$ &   $P$      &  $P^2$    &   $E$\\
$P^3Q$   & $P^3Q$     &      $PQ$    &   $Q$      &    $P^2Q$&   $P$   &   $P^3$    &  $E$      &   $P^2$
\end{tabular}
\end{ruledtabular}
\end{table}

\begin{table}[htbp]
\caption{Multiplication table of $R_2(D_2)$. Notice that $P^4=Q^2=E$
and $QP=P^3Q$.} \label{tab:R2}
\begin{ruledtabular}
\begin{tabular}{c|cccccccc}
         & $E$        &      $P^2$   & $P^3$      &    $P$   &    $Q$  &    $P^2Q$  &   $PQ$    &   $P^3Q$\\
\hline
$E$      & $E$        &      $P^2$   &   $P^3$    &    $P$   &    $Q$  &    $P^2Q$  &  $PQ$     &   $P^3Q$\\
$P^2$    & $P^2$      &      $E$     &   $P$      &    $P^3$ &   $P^2Q$&    $Q$     &  $P^3Q$   &   $PQ$ \\
$P^3$    & $P^3$      &      $P$     &   $P^2$    &    $E$   &  $P^3Q$ &    $PQ$    &  $Q$      &   $P^2Q$\\
$P$      & $P$        &      $P^3$   &    $E$     &    $P^2$ &   $PQ$  &    $P^3Q$  &  $P^2Q$   &   $Q$  \\
$Q$      & $Q$        &      $P^2Q$  &    $PQ$    &    $P^3Q$&   $E$   &    $P^2$   &  $P^3$    &   $P$ \\
$P^2Q$   & $P^2Q$     &      $Q$     &   $P^3Q$   &    $PQ$  &   $P^2$ &    $E$     &  $P$      &   $P^3$  \\
$PQ$     & $PQ$       &      $P^3Q$  &   $P^2Q$   &    $Q$   &   $P$   &    $P^3$   &  $E$      &   $P^2$\\
$P^3Q$   & $P^3Q$     &      $PQ$    &   $Q$      &    $P^2Q$&   $P^3$ &    $P$     &  $P^2$    &   $E$
\end{tabular}
\end{ruledtabular}
\end{table}
To obtain all the irreducible representations, we only need to block
diagonalize the canonical representation matrices of the two
generators $P$ and $Q$. In the canonical representation, the group
space itself is also the representation space. Each group element
$g_1$ is considered as an operator $\hat g_1$:
\begin{eqnarray}\label{canonical}
\hat g_1(g_2)=g_1g_2.
\end{eqnarray}
here $g_2$ and $g_1g_2$ are two vectors in the representation space
and $\hat g_1$ becomes a matrix.

 The
canonical representation matrices of the generators of $R_1$ can be
read from table \ref{tab:Rl}:
\begin{eqnarray*}
P=\left(\begin{matrix}
       0&0&1&0&0&0&0&0\\
       0&0&0&1&0&0&0&0\\
       0&1&0&0&0&0&0&0\\
       1&0&0&0&0&0&0&0\\
       0&0&0&0&0&0&0&1\\
       0&0&0&0&0&0&1&0\\
       0&0&0&0&1&0&0&0\\
       0&0&0&0&0&1&0&0
\end{matrix}\right),
Q=\left(\begin{matrix}
       0&0&0&0&0&1&0&0\\
       0&0&0&0&1&0&0&0\\
       0&0&0&0&0&0&0&1\\
       0&0&0&0&0&0&1&0\\
       1&0&0&0&0&0&0&0\\
       0&1&0&0&0&0&0&0\\
       0&0&1&0&0&0&0&0\\
       0&0&0&1&0&0&0&0
\end{matrix}\right),
\end{eqnarray*}
To simultaneously block diagonalize the above matrices, we need to
identify the base vectors (or wave function) of each irreducible
representation (these base vectors form a unitary matrix which
block diagonalize $P$ and $Q$ simultaneously). In quantum mechanics,
we use good quantum numbers (eigenvalues of commuting quantities) to
label different states. For example, $|S,m\rangle$ symbols a spin
state, where $S(S+1)$ is the eigenvalue of the Casimir operator of
 $SO(3)$ group and $m$ the eigenvalue of the Casimir operator of
its subgroup $SO(2)$. Similar method has been applied to
the representation theory of groups.\cite{ChenJQ} What we need to do is
to find all the commuting quantities, or the complete set of
commuting operators (CSCO).\cite{ChenJQ}

The Casimir operators of discrete groups are their class operators.
For $R_1(D_2)$, there are five classes (hence there are five
different irreducible linear representations), and the corresponding
five class operators are given as below:
\begin{equation*}
C=\{{E},\ {P^2},\ {P+P^3},\ {Q+P^2Q},\ {PQ+P^3Q}\}.
\end{equation*}
The class operators commute with each other and all other group elements.
This set of class operators $C$ is
called CSCO-I in Ref.~\onlinecite{ChenJQ}. The eigenvalues of the
class operators are greatly degenerate, which can only be used to
distinguish different irreducible representations(IRs). To
distinguish the bases in each IR, we can use the class operators of
its subgroup(s). Group $R_1$ has a cyclic subgroup
\[C(s)=\{E,P,P^2,P^3\},\]
each element forms a class. The set of class operators of the
subgroup is written as $C(s)$. The operator-set $(C,C(s))$ is called
CSCO-II, which can be used to distinguish all the bases if every IR
occurs only once in the reduced canonical representation.

However, in the reduced canonical representation, a $d$-dimensional
representation occurs $d$ times and they have the same eigenvalues
for CSCO-II. To lift this degeneracy, we need more commuting
operators. Fortunately, we can use the class operators of the
`intrinsic group' $\bar R_1$, whose group elements are defined as
follows
\begin{eqnarray}
\hat {\bar g}_1(g_2)=g_2g_1.
\end{eqnarray}
Notice that $\hat {\bar g}$ commutes with $\hat {g}$ defined in
(\ref{canonical}). The class operators of $\bar R_1$ are identical
to those of $R_1$, $\bar C=C$. The set of class operators for the
intrinsic subgroup
\[\bar C(s)=\{\bar E, \bar P, \bar P^2,\bar P^3\}\]
is noted as $\bar C(s)$. The eigenvalues of $\bar C(s)$ provide a
different set of `quantum numbers' to each identical IR.

Now we obtain the complete set of class operators $\left(C,C(s),\bar
C(s)\right)$, which is called CSCO-III. The common eigenvectors of
the operators in CSCO-III are the orthonormal bases of the
irreducible representations, and each eigenvector has a unique
`quantum number'.

To obtain the bases, we need to simultaneously diagonalize all the
operators in CSCO-III and get their eigenvectors. Actually, we only
need a few of these operators, for example, we can choose $Q+P^2Q$
in $C$, $P$ in $C(s)$ and $\bar P$ in $\bar C(s)$. The matrices of
these operators of $R_1(D_2)$ are given below:
\begin{eqnarray*}
&&Q+P^2Q=\left(\begin{matrix}
       0&0&0&0&1&1&0&0\\
       0&0&0&0&1&1&0&0\\
       0&0&0&0&0&0&1&1\\
       0&0&0&0&0&0&1&1\\
       1&1&0&0&0&0&0&0\\
       1&1&0&0&0&0&0&0\\
       0&0&1&1&0&0&0&0\\
       0&0&1&1&0&0&0&0
\end{matrix}\right),\\
&&P=\left(\begin{matrix}
       0&0&1&0&0&0&0&0\\
       0&0&0&1&0&0&0&0\\
       0&1&0&0&0&0&0&0\\
       1&0&0&0&0&0&0&0\\
       0&0&0&0&0&0&0&1\\
       0&0&0&0&0&0&1&0\\
       0&0&0&0&1&0&0&0\\
       0&0&0&0&0&1&0&0
\end{matrix}\right),\\
&&\bar P=\left(\begin{matrix}
       0&0&1&0&0&0&0&0\\
       0&0&0&1&0&0&0&0\\
       0&1&0&0&0&0&0&0\\
       1&0&0&0&0&0&0&0\\
       0&0&0&0&0&0&1&0\\
       0&0&0&0&0&0&0&1\\
       0&0&0&0&0&1&0&0\\
       0&0&0&0&1&0&0&0
\end{matrix}\right),
\end{eqnarray*}

Practically, we can diagonalize a linear combination $\hat
O=(Q+P^2Q)+aP+b\bar P$, where $a,b$ are arbitrary constants ensuring
that all the eigenvalues of $\hat O$ are non-degenerate. From the
non-degenerate eigenvectors (column vectors) of $\hat O$, we obtain
an unitary matrix $U$:
\begin{eqnarray}\label{U1}
&U=\left(\begin{matrix}
   {1\over\sqrt8}& {1\over\sqrt8}&  {1\over\sqrt8}& {1\over\sqrt8}& {1\over2}    &     0       & 0         &   {1\over2}\\
   {1\over\sqrt8}& {1\over\sqrt8}&  {1\over\sqrt8}& {1\over\sqrt8}&-{1\over2}    &     0       & 0         &  -{1\over2}\\
   {1\over\sqrt8}& {1\over\sqrt8}& -{1\over\sqrt8}&-{1\over\sqrt8}&  -{1\over2}i &     0       & 0         &  {1\over2}i\\
   {1\over\sqrt8}& {1\over\sqrt8}& -{1\over\sqrt8}&-{1\over\sqrt8}& {1\over2}i   &     0       & 0         & -{1\over2}i\\
   {1\over\sqrt8}&-{1\over\sqrt8}&  {1\over\sqrt8}&-{1\over\sqrt8}&  0           & -{1\over2}i &-{1\over2}i&           0\\
   {1\over\sqrt8}&-{1\over\sqrt8}&  {1\over\sqrt8}&-{1\over\sqrt8}&  0           & {1\over2}i  & {1\over2}i&           0\\
   {1\over\sqrt8}&-{1\over\sqrt8}& -{1\over\sqrt8}& {1\over\sqrt8}&  0           &  -{1\over2} & {1\over2} &           0\\
   {1\over\sqrt8}&-{1\over\sqrt8}& -{1\over\sqrt8}& {1\over\sqrt8}&  0           &  {1\over2}  &-{1\over2} &           0
\end{matrix}\right),\nonumber\\
\end{eqnarray}
The matrix $U$ is the transformation that block diagonalizes $P$ and
$Q$ simultaneously:
\begin{eqnarray}\label{DigR1}
&&U^\dagger PU=\left(\begin{matrix}
       1&0& 0&0&0& 0& 0&0\\
       0&1& 0&0&0& 0& 0&0\\
       0&0&-1&0&0& 0& 0&0\\
       0&0&0&-1&0& 0& 0&0\\
       0&0& 0&0&-i&0& 0&0\\
       0&0& 0&0&0& i& 0&0\\
       0&0& 0&0&0& 0&-i&0\\
       0&0& 0&0&0& 0& 0&i
\end{matrix}\right),\\
&&U^\dagger QU=\left(\begin{matrix}
       1&0&0& 0&0&0&0&0\\
       0&-1&0&0&0&0&0&0\\
       0&0&1& 0&0&0&0&0\\
       0&0&0&-1&0&0&0&0\\
       0&0&0& 0&0&i&0&0\\
       0&0&0& 0&i&0&0&0\\
       0&0&0& 0&0&0&0&i\\
       0&0&0& 0&0&0&i&0
\end{matrix}\right),
\end{eqnarray}
There are four 1-d IRs and one 2-d IR (which occurs twice) in the
reduced canonical representation:
\begin{eqnarray}\label{R1_2d}
P_2=\left(\begin{matrix}
       -i& 0\\
       0 & i
\end{matrix}\right)=-i\sigma_z,
Q_2=\left(\begin{matrix}
       0&i\\
       i&0
\end{matrix}\right)=i\sigma_x,
\end{eqnarray}
So there are totally five independent representations.

The eight elements in $R_1$ can be projected onto $D_2$, as shown in
Tab.~\ref{tab:prj1}. And the linear representations of $R_1$
correspond to the projective representations of $D_2$. The four 1-d
IRs correspond to the linear IRs of $D_2$, and the 2-d IR stands for
a nontrivial projective IR of $D_2$. Up to a phase factor, these 2-d
matrices are the $180^\circ$ rotation operators of a spin with
$J=1/2$ (which is a projective IR of SO(3) group).
\begin{table}[htbp]
\caption{Projection from $R_1(D_2$) to $D_2$ } \label{tab:prj1}
\begin{ruledtabular}
\begin{tabular}{c|cccc}
$R_1(D_2)$          & $E$         &    $P$    &    $Q$    &   $PQ$     \\
                    &  $P^2$      &    $P^3$  &  $P^2Q$   &   $P^3Q$   \\
\hline%
$D_2$               &   $E$       &    $R_z$  & $R_x$     & $R_y$      \\%
\hline%
rotation $\pi$ of $J=1/2$\\(up to a phase factor) &    $I$      &$i\sigma_z$&$i\sigma_x$&$i\sigma_y$%
\end{tabular}
\end{ruledtabular}
\end{table}

Now let's look at the direct product of the projective IRs of $D_2$. For
the 1-d linear IRs, the direct product are still 1-d IRs, which
satisfy the following law:
\begin{eqnarray*}
&&A\times B_1=B_1,\ A\times B_2=B_2,\ A\times B_3=B_3,\\
&&B_1\times B_2=B_3,\ B_1\times B_3=B_2,\ B_2\times B_3=B_1.\
\end{eqnarray*}
The direct product of 1-d and 2-d IRs are still 2-d projective IRs
of $D_2$. The direct product of two 2-d projective IRs is
interesting. It reduces to four 1-d linear IRs. Using the CSCO-II,
we can diagonalize the $4\times4$ matrices $P_2\otimes P_2$ and
$Q_2\otimes Q_2$ with the following unitary matrix (the column
vectors are just the CG coefficients):
\begin{eqnarray*}
U_4=\left(\begin{matrix}
-{1\over\sqrt2}& {i\over\sqrt2} &             0&0\\
              0&               0&{1\over\sqrt2}& {1\over\sqrt2}\\
              0&               0&{1\over\sqrt2}&-{1\over\sqrt2}\\
 {1\over\sqrt2}& {i\over\sqrt2} &             0&0
\end{matrix}\right),\\
U_4^\dagger (P_2\otimes P_2)U_4=\left(\begin{matrix}
      -1& 0& 0&0\\
       0&-1& 0&0\\
       0& 0& 1&0\\
       0& 0& 0&1
\end{matrix}\right),\\
U_4^\dagger (Q_2\otimes Q_2)U_4=\left(\begin{matrix}
       1&0& 0& 0\\
       0&-1& 0& 0\\
       0&0&-1& 0\\
       0&0& 0& 1
\end{matrix}\right),
\end{eqnarray*}
These CG coefficients of the projective IRs of $D_2$ are analogous
to the decoupling of the direct product of two spins with $J=1/2$,
${1\over2}\otimes{1\over2} =1\otimes0$ except the 3-d IR of spin-1
becomes a direct sum of three 1-d IRS. If we label the bases of the
2-d projective representation of $D_2$ as $|\uparrow\rangle,
|\downarrow\rangle$, then the CG coefficients are given as:
\begin{eqnarray}\label{CG}
&&|x\rangle={1\over\sqrt2}(|\downarrow_1\downarrow_2\rangle-|\uparrow_1\uparrow_2\rangle),\nonumber\\
&&|y\rangle={i\over\sqrt2}(|\downarrow_1\downarrow_2\rangle+|\uparrow_1\uparrow_2\rangle),\nonumber\\
&&|z\rangle={1\over\sqrt2}(|\uparrow_1\downarrow_2\rangle+|\downarrow_1\uparrow_2\rangle),\nonumber\\
&&|\mathrm{singlet}\rangle={1\over\sqrt2}(|\uparrow_1\downarrow_2\rangle-|\downarrow_1\uparrow_2\rangle),
\end{eqnarray}

Repeating the above procedure, we obtain the IRs of $R_2(D_2)$. The four
1-d IRs are the same as that of $R_1(D_2)$, while the 2-d IR is
given as:
\begin{eqnarray}\label{R2_2d}
P_2'=\left(\begin{matrix}
       -i& 0\\
       0 & i
\end{matrix}\right)=-i\sigma_z,
Q_2'=\left(\begin{matrix}
       0&1\\
       1&0
\end{matrix}\right)=\sigma_x,\nonumber\\
\end{eqnarray}
The above representation and Eq.~(\ref{R1_2d}) differ only by a
gauge transformation $P_2'=P_2$ and $Q_2'=iQ_2$, so they belong to
the same projective representation of $D_2$. The CG coefficients for
the 2-d IRs are obtained easily:
\begin{eqnarray}\label{CG2}
&&|x\rangle={1\over\sqrt2}(|\downarrow_1\downarrow_2\rangle+|\uparrow_1\uparrow_2\rangle),\nonumber\\
&&|y\rangle={i\over\sqrt2}(|\downarrow_1\downarrow_2\rangle-|\uparrow_1\uparrow_2\rangle),\nonumber\\
&&|z\rangle={1\over\sqrt2}(|\uparrow_1\downarrow_2\rangle-|\downarrow_1\uparrow_2\rangle),\nonumber\\
&&|\mathrm{singlet}\rangle={1\over\sqrt2}(|\uparrow_1\downarrow_2\rangle+|\downarrow_1\uparrow_2\rangle).
\end{eqnarray}

\subsection{sMP state with $D_2$ symmetry and its parent Hamiltonian}

Before studying the model with $D_2$ symmetry, let's review the
$S=1$ AKLT model\cite{AKLT} (which has $SO(3)$ symmetry) first. The
AKLT state is a sMP state given by $A^x=\sigma_x, A^y=\sigma_y,
A^{z}=\sigma_z$. The $A^m$ matrices are two by two, meaning that the
physical spin $S=1$ is viewed as symmetric combination of two
$J=1/2$ virtual spins (essentially projective representations of
$SO(3)$). Alternatively, we can write the state as
\begin{eqnarray}
|\phi\rangle=\mathrm{Tr}(W_1W_2...W_N),
\end{eqnarray}
where $W_i=A^x|x\rangle_i+A^y|y\rangle_i+A^z|z\rangle_i$. According
to Ref.~\onlinecite{Tu}, the matrices $A^m$ of a sMP state can be
obtained by
\begin{eqnarray}\label{A}
A^m=B^T(C^m)^*,
\end{eqnarray}
where $B$ is the CG coefficient combining two virtual spins into a
singlet $|\mathrm{0,0}\rangle= B_{m_1m_2}|{1\over2},m_1;{1\over2},
m_2\rangle$, and $C^m$ is the CG coefficient combining two virtual
spins into a triplet $|1,m\rangle=C^m_{m_1m_2}|{1\over2},m_1;
{1\over2}, m_2 \rangle$.

Now we can generalize this formalism to the $D_2$ case, where the
three states of $S=1$ become a direct sum of three IRs of $D_2$.
$D_2$ group has a 2-dimensional nontrivial projective
representation, and the direct product of two such projective IRs
can be reduced using the CG coefficients ($B$ and $C^{x,y,z}$) in
Eq.~(\ref{CG}). Similar to the $SO(3)$ case, we can consider the two
2-d projective IRs as `virtual spins'. From eq.~(\ref{A}), we can
construct the following matrix (similar sMP state has been studied
in Ref.~\onlinecite{D2})
\begin{eqnarray}\label{D2MPS}
W=a\sigma_x|x\rangle + b\sigma_y|y\rangle + c\sigma_z|z\rangle,
\end{eqnarray}
where $a,b,c$ are arbitrary nonzero \textit{complex} constants. The
corresponding sMP state is given by $|\phi\rangle=\mathrm{Tr}
(W_1W_2...W_N)$, which is invariant under the group $D_2$. Notice
that the CG coefficients in Eq.~(\ref{CG2}) give the same sMP state
(up to some gauge transformations). Notice also that (\ref{D2MPS})
is different from (\ref{T0}), (\ref{Tx}), (\ref{Ty}) or (\ref{Tz}).
If $a, b$ or $c$ are to be arbitrary complex numbers, it is not
invariant under $T$.

The above sMP state is injective, and the parent Hamiltonian can be
obtained by projection operators. We consider a block containing two
spins, the four matrix elements of $W_iW_{i+1}$ span a 4-dimensional
Hilbert space. Suppose the orthonormal bases are
$|\psi_{1,2,3,4}\rangle_i $, then we can construct a projector
\begin{eqnarray}
P_i=1-\sum_{\alpha=1}^4|\psi_\alpha\rangle\langle\psi_\alpha|_i,
\end{eqnarray}
and the Hamiltonian $H=\sum_iP_i$. It can be easily checked that the
sMP state is the unique ground state of this Hamiltonian.

The projector $P_i$ is a nine by nine matrix that can be written
in forms of spin operators. Notice that any Hermitian operator of
site $i,j$ can be expanded by the 81 generators of
$U(9)=U(3)_i\otimes U(3)_{j}$, i.e., $\lambda_{\alpha
i}\lambda_{\beta j}$ ($\alpha,\beta=1,...,9$). So, we have
\begin{eqnarray}
P_i=\sum_{\alpha,\beta=1}^9\xi_{\alpha\beta}\lambda_{\alpha
i}\lambda_{\beta j},
\end{eqnarray}
where $\xi_{\alpha\beta}$ are constants. Further, the generators of
$U(3)$ can be written as polynomials of spin operators.
\begin{eqnarray*}
&&\lambda_1=(S_x+S_{xz})/\sqrt2,\\
&&\lambda_6=(S_x-S_{xz})/\sqrt2,\\
&&\lambda_2=(S_y+S_{yz})/\sqrt2,\\
&&\lambda_7=(S_y-S_{yz})/\sqrt2,\\
&&\lambda_4=S_x^2-S_y^2,\\
&&\lambda_5=S_{xy},\\
&&\lambda_3=(S_z+3S_z^2)/2-I,\\
&&\lambda_8=(3S_z-3S_z^2+2I)/2\sqrt3,\\
&&\lambda_9=\sqrt{2\over3}I=\sqrt{1\over6}(S_x^2+S_y^2+S_z^2).
\end{eqnarray*}
where $S_{mn}=S_m S_n+S_n S_m, (m,n=x,y,z)$ and
$\lambda_1\sim\lambda_8$ are the Gellmann matrices of $SU(3)$
generators. Finally, we can write the Hamiltonian in forms of spin
operators. For simplicity, we first assume $a,b,c$ are real numbers,
then the Hamiltonian is given in (\ref{Ht0}), which is invariant
under $T$. The $T$ symmetry goes away when $a,b$ or $c$ becomes an
arbitrary complex number. For instance, if $a\to ae^{i\theta}$, then
the Hamiltonian (\ref{Ht0}) becomes
\begin{widetext}
\begin{eqnarray}\label{H_mps2}
H&=&\sum_i\left[({1\over4}+{b^2c^2/2\over a^4 + b^4 + c^4
})S_{x,i}S_{x,j} + ({1\over4}+{\cos2\theta a^2c^2/2\over a^4 + b^4 +
c^4 })S_{y,i}S_{y,j} -{\sin2\theta {a^2c^2}/2\over
a^4 + b^4 +c^4}(S_{y,i}S_{xz,j}+S_{xz,i}S_{y,j})\right. \nonumber\\
&&\left.+  ({1\over4}+{\cos2\theta a^2b^2/2\over a^4 + b^4 + c^4
})S_{z,i}S_{z,j}+ {\sin2\theta  a^2b^2/2\over a^4 + b^4 + c^4
}(S_{z,i}S_{xy,j}+S_{xy,i}S_{z,j})+ ({1\over4}-{\cos2\theta
a^2b^2/2\over a^4 + b^4 + c^4 })S_{xy,i}S_{xy,j}\right. \nonumber\\
&&\left. + ({1\over4}-{b^2c^2/2\over a^4 + b^4 + c^4
})S_{yz,i}S_{yz,j} +
({1\over4}-{\cos2\theta a^2c^2/2\over a^4 + b^4 + c^4 })S_{xz,i}S_{xz,j}\right] +h_0
\end{eqnarray}
\end{widetext}
When $\sin2\theta\neq0$ above Hamiltonian does not
have $T$ symmetry.

Varying the values of $a,b,c$, we can transform the ground state of
the above Hamiltonian into that of the AKLT model smoothly without
breaking $D_2$ symmetry. This means that above sMP state also
belongs to the Haldane phase. In appendix \ref{appB} we will
consider the models with additional time reversal symmetry.

\section{Spin Chain with $D_{2h}$ symmetry}\label{appB}

In the last section we have studied the spin chain with on-site
$D_2$ symmetry. Now we consider a $S=1$ spin chain with additional
spin-inversion (or time-reversal) symmetry. The complete on-site
symmetry now becomes $D_{2h}=\{E, R_x, R_y, R_z,T,R_xT,R_yT,R_zT\}$.
It has eight 1-d linear real IRs, as listed in Tab.~\ref
{tab:Repd2h}. Notice the time reversal operator $T=e^{-i\pi S_y}K$
is anti-unitary, so the states $|m\rangle$ and $i|m\rangle$
($m=x,y,z$) belong to different linear representations, the former
is odd under $T$ and is noted by index $u$, and the latter is even
under $T$ as noted by $g$. So we need to introduce six bases
$|x\rangle, |y\rangle, |z\rangle$ and $i|x\rangle, i|y\rangle,
i|z\rangle$. To construct a sMP state, at least one of the pair
$|x\rangle$, $i|x\rangle$ (and also the pairs $|y\rangle$,
$i|y\rangle$ and $|z\rangle$, $i|z\rangle$) should be present in the
physical bases.
\begin{table}[htbp]
\caption{Linear representations of $D_{2h}$} \label{tab:Repd2h}
\begin{ruledtabular}
\begin{tabular}{c|cccccccc|ccc}
            & $E$        & $R_x$        & $R_y$      & $R_z$    & $T$ & $R_xT$       & $R_yT$     & $R_zT$&bases&operators&\\
\hline
$A_{g}$     &      1     &       1      &      1     &     1    &  1  &       1      &      1     &     1    &$|0,0\rangle$&$S_x^2,S_y^2,S_z^2$& \\
$B_{1g}$    &      1     &      -1      &     -1     &     1    &  1  &      -1      &     -1     &     1    &$i|1,z\rangle$&$S_{xy}$& \\
$B_{2g}$    &      1     &      -1      &      1     &    -1    &  1  &      -1      &      1     &    -1    &$i|1,y\rangle$&$S_{xz}$& \\
$B_{3g}$    &      1     &       1      &     -1     &    -1    &  1  &       1      &     -1     &    -1    &$i|1,x\rangle$&$S_{yz}$& \\
\hline
$A_{u}$     &      1     &       1      &      1     &     1    & -1  &      -1      &     -1     &    -1    &$i|0,0\rangle$&   &\\
$B_{1u}$    &      1     &      -1      &     -1     &     1    & -1  &       1      &      1     &    -1    &$|1,z\rangle$&$S_z$&\\
$B_{2u}$    &      1     &      -1      &      1     &    -1    & -1  &       1      &     -1     &     1    &$|1,y\rangle$&$S_y$& \\
$B_{3u}$    &      1     &       1      &     -1     &    -1    & -1  &      -1      &      1     &     1    &$|1,x\rangle$&$S_x$& 
\end{tabular}
\end{ruledtabular}
\end{table}

To obtain the projective IRs of $D_{2h}$, we need to study the
linear IRs of the representation group $R(D_{2h})$, which also has
three generators $P,Q,R$ (corresponding to $R_z, R_x, T$) satisfying
$P^4=Q^4=R^4=E$ and $P^3Q=QP,Q^3R=RQ,R^3P=PR$.\cite{Boyal} The total
number of elements in $R(D_{2h})$ is 64. It has 8 1-d
representations (corresponding to the 8 linear IRs of $D_{2h}$) and
14 2-d representations (corresponding to the 7 classes of projective
IRs of $D_{2h}$). To obtain the IRs of $R(D_{2h})$, we only need to
know the representation matrix of the three generators $P,Q,R$.
Using the same method given in the last section, we obtain all the IRs
of $R(D_{2h})$ (see Tab.~\ref{tab:prjRep}).

\begin{table*}[htbp]
\caption{Projective representations of $D_{2h}$. The numbers
$\alpha, \beta,\gamma$ are obtained by $\alpha=P^2,\ \beta=Q^2,\
\gamma=R^2$. The three generators $P,Q,R$ of $R(D_{2h})$ will
project to $R_z,R_x,T$ of $D_{2h}$, respectively. }
\label{tab:prjRep}
\begin{ruledtabular}
\begin{tabular}{|c|cccc|c|}
&$P(R_z)$&$Q(R_x)$&$R(T)$&...&$\alpha=P^2,\ \beta=Q^2,\ \gamma=R^2$\\
\hline
$A_{g}$     &     1      &      1     &      1     &...&  \\
$B_{1g}$    &     1      &     -1     &      1     &...&  \\
$B_{2g}$    &    -1      &     -1     &      1     &...&  \\
$B_{3g}$    &    -1      &      1     &      1     &...& 1 1 1  \\
$A_{u}$     &     1      &      1     &     -1     &...&  \\
$B_{1u}$    &     1      &     -1     &     -1     &...&  \\
$B_{2u}$    &    -1      &     -1     &     -1     &...&  \\
$B_{3u}$    &    -1      &      1     &     -1     &...&  \\
\hline
$E_1$       &     I      & $i\sigma_z$& $ \sigma_y$&...& 1 -1 1 \\
$E_2=E_1\otimes B_{3g}$       &    -I      & $i\sigma_z$& $ \sigma_y$&...&  \\
\hline
$E_3$       & $ \sigma_z$&     I      & $i\sigma_y$&...& 1 1 -1 \\
$E_4=E_3\otimes B_{1g}$       & $ \sigma_z$&    -I      & $i\sigma_y$&...& \\
\hline
$E_5$       & $i\sigma_z$& $ \sigma_x$&     I      &...& -1 1 1 \\
$E_6=E_5\otimes A_{u}$       & $i\sigma_z$& $ \sigma_x$&    -I      &...& \\
\hline
$E_7$       & $ \sigma_z$& $i\sigma_z$& $i\sigma_x$&...&1 -1 -1 \\
$E_8=E_7\otimes B_{1g}$       & $ \sigma_z$&-$i\sigma_z$& $i\sigma_x$&...& \\
\hline
$E_9$       & $i\sigma_z$& $ \sigma_x$& $i\sigma_x$&...&-1 1 -1 \\
$E_{10}=E_9\otimes A_{u}$    & $i\sigma_z$& $ \sigma_x$&-$i\sigma_x$&...& \\
\hline
$E_{11}$    & $i\sigma_z$& $i\sigma_x$& $ \sigma_z$&...&-1 -1 1 \\
$E_{12}=E_{11}\otimes B_{3g}$    &$i\sigma_z$& $i\sigma_x$& -$ \sigma_z$&...& \\
\hline
$E_{13}$    & $i\sigma_z$& $i\sigma_x$& $i\sigma_y$&...&-1 -1 -1 \\
$E_{14}=E_{13}\otimes A_{u}$    & $i\sigma_z$& $i\sigma_x$&-$i\sigma_y$&...& \\
\end{tabular}
\end{ruledtabular}
\end{table*}

Now we give the CG coefficients that reduce the direct product of
two projective IRs into direct sum of linear IRs of $D_{2h}$.

\begin{widetext}
\begin{eqnarray}\label{CGd2h_1}
&&E_1\otimes E_1=E_2\otimes E_2=A_g\oplus B_{1g}\oplus A_u\oplus B_{1u};\ \ \ C^{A_g}=\sigma_x,\ C^{B_{1g}}=\sigma_z,\ C^{A_u}=i\sigma_y,\ C^{B_{1u}}=I;\nonumber\\
&&E_3\otimes E_3=E_4\otimes E_4=A_g\oplus B_{3g}\oplus A_u\oplus B_{3u};\ \ \ C^{A_g}=I,\ C^{B_{3g}}=i\sigma_y,\ C^{A_u}=\sigma_z,\ C^{B_{3u}}=\sigma_x;\nonumber\\
&&E_5\otimes E_5=E_6\otimes E_6=A_g\oplus B_{1g}\oplus B_{2g}\oplus B_{3g};\ \ \ C^{A_g}=\sigma_x,\ C^{B_{1g}}=i\sigma_y,\ C^{B_{2g}}=\sigma_z,\ C^{B_{3g}}=I;\nonumber\\
&&E_7\otimes E_7=E_8\otimes E_8=B_{1g}\oplus B_{3g}\oplus B_{1u}\oplus B_{3u};\ \ \ C^{B_{1g}}=\sigma_z,\ C^{B_{3g}}=i\sigma_y,\ C^{B_{1u}}=I,\ C^{B_{3u}}=\sigma_x;\nonumber\\
&&E_9\otimes E_9=E_{10}\otimes E_{10}=B_{1g}\oplus B_{2g}\oplus A_u\oplus B_{3u};\ \ \ C^{A_u}=\sigma_x,\ C^{B_{3u}}=I,\ C^{B_{1g}}=i\sigma_y,\ C^{B_{2g}}=\sigma_z;\nonumber\\
&&E_{11}\otimes E_{11}=E_{12}\otimes E_{12}=B_{2g}\oplus B_{3g}\oplus A_u\oplus B_{1u};\ \ \ C^{A_u}=i\sigma_y,\ C^{B_{1u}}=\sigma_x,\ C^{B_{2g}}=I,\ C^{B_{3g}}=\sigma_z;\nonumber\\
&&E_{13}\otimes E_{13}=E_{14}\otimes E_{14}=A_{g}\oplus B_{2g}\oplus B_{1u}\oplus B_{3u};\ \ \ C^{A_{g}}=i\sigma_y,\ C^{B_{2g}}=I,\ C^{B_{1u}}=\sigma_x,\ C^{B_{3u}}=\sigma_z;%
\end{eqnarray}
and
\begin{eqnarray}\label{CGd2h_2}
&&E_1\otimes E_2=B_{2g}\oplus B_{3g}\oplus B_{2u}\oplus B_{3u};\ \ \ C^{B_{3g}}=\sigma_x,\ C^{B_{2g}}=\sigma_z,\ C^{B_{3u}}=i\sigma_y,\ C^{B_{2u}}=I;\nonumber\\
&&E_3\otimes E_4=B_{1g}\oplus B_{2g}\oplus B_{1u}\oplus B_{2u};\ \ \ C^{B_{1g}}=I,\ C^{B_{2g}}=i\sigma_y,\ C^{B_{1u}}=\sigma_z,\ C^{B_{2u}}=\sigma_x;\nonumber\\
&&E_5\otimes E_6=A_u\oplus B_{1u}\oplus B_{2u}\oplus B_{3u};\ \ \ C^{A_u}=\sigma_x,\ C^{B_{1u}}=i\sigma_y,\ C^{B_{2u}}=\sigma_z,\ C^{B_{3u}}=I;\nonumber\\
&&E_7\otimes E_8=A_{g}\oplus B_{2g}\oplus A_{u}\oplus B_{2u};\ \ \ C^{A_{g}}=\sigma_z,\ C^{B_{2g}}=i\sigma_y,\ C^{A_{u}}=I,\ C^{B_{2u}}=\sigma_x;\nonumber\\
&&E_9\otimes E_{10}=A_{g}\oplus B_{3g}\oplus B_{1u}\oplus B_{2u};\ \ \ C^{A_{g}}=\sigma_x,\ C^{B_{3g}}=I,\ C^{B_{1u}}=i\sigma_y,\ C^{B_{2u}}=\sigma_z;\nonumber\\
&&E_{11}\otimes E_{12}=A_{g}\oplus B_{1g}\oplus B_{2u}\oplus B_{3u};\ \ \ C^{A_{g}}=i\sigma_y,\ C^{B_{1g}}=\sigma_x,\ C^{B_{2u}}=I,\ C^{B_{3u}}=\sigma_z;\nonumber\\
&&E_{13}\otimes E_{14}=B_{1g}\oplus B_{3g}\oplus A_{u}\oplus B_{2u};\ \ \ C^{A_{u}}=i\sigma_y,\ C^{B_{2u}}=I,\ C^{B_{1g}}=\sigma_x,\ C^{B_{3g}}=\sigma_z.%
\end{eqnarray}
\end{widetext}
Here all the coefficients are chosen to be real.

Now we construct sMP states from the CG coefficients
Eqs.~(\ref{CGd2h_1}), (\ref{CGd2h_2}) and (\ref{A}). Since all the
CG coefficients are real, the constructed matrices $A^m=B^T(C^m)^*$
are also real (here $B=C^{A_g}$, $m=B_{1g},B_{1u}...B_{3u}$), and
are invariant under the anti-unitary operator $K$. However, the
bases $|B_{1g}\rangle=i|z\rangle, |B_{2g}\rangle=i|y\rangle$ or
$|B_{3g}\rangle=i|z\rangle$ contain a factor $i$, this factor $i$
may be combined with $A^m$ when writing the matrix $W=\sum_mA^m
|m\rangle$. So the definition of $A^m$ depends on the choice of
base. If we choose $|m\rangle=|x\rangle, |y\rangle, |z\rangle$ as
the physical bases, then $A^m$ will absorb the factor $i$ (if
existent) and may be either real or purely imaginary. This
convention is adopted in the main part of this paper. On the other
hand, if we just choose $|m\rangle=|B_{1g}\rangle, |B_{1u}\rangle,
... |B_{3u}\rangle$ as the physical bases (and forget about the
factor that some bases, such as $B_{1g}$ and $B_{1u}$, are linearly
dependent), then all the matrices $A^m$ are real. In the following
discussion, we will adopt the second convention.

Notice that the combinations $E_{5}\otimes E_{5}$, $E_{9}\otimes
E_{10}$, $E_{11}\otimes E_{12}$, $E_{13}\otimes E_{13}$ contain all
the bases of $S=1$ ($|B_1\rangle, |B_2\rangle, |B_3\rangle$) and the
singlet state ($|A_g\rangle$), we can construct sMP state using
these combinations. We will study them case by case.

1)$E_5\otimes E_5$\\
Up to an overall phase, the local matrix $W$ is given by
$W=a\sigma_x|x\rangle +ib\sigma_y|y\rangle+c\sigma_z|z\rangle$, here
$a,b,c$ are real numbers. The Hamiltonian can be constructed using
the method given in appendix {\ref{appA}}, and the result is given
in (\ref{Hty}).

2)$E_9\otimes E_{10}$\\
Up to an overall phase, the local matrix $W$ is given by
$W=a\sigma_x|x\rangle + b\sigma_y|y\rangle + ic\sigma_z|z\rangle$,
and the Hamiltonian is shown in (\ref{Htz}).

3)$E_{11}\otimes E_{12}$\\
Up to an overall phase, the local matrix $W$ is given by
$W=ia\sigma_x|x\rangle + b\sigma_y|y\rangle + c\sigma_z|z\rangle$,
and the Hamiltonian is given in (\ref{Htx}).

4)$E_{13}\otimes E_{13}$\\
The local matrix $W$ is given by $W=a\sigma_x|x\rangle +
b\sigma_y|y\rangle + c\sigma_z|z\rangle$, and the Hamiltonian is
given in (\ref{Ht0}).

With the $D_{2h}$ symmetry kept, the ground states of the above four
exactly solvable models cannot be smoothly transformed into each
other, which indicates they belong to different SPT phases (see
section III).

According to Ref.~\onlinecite{CGW2}, there should be seven SPT
phases since there are seven classes of projective representations.

However, in the other three projective IRs, the reduced Hilbert
space of the direct product of two virtual `spins' only contains one
of the three bases for the physical $S=1$ states (notice that the
singlet $|A_g\rangle$ is necessary to construct a sMP state), which
means that these three SPT phases cannot be realized in $S=1$
systems.

\section{Invariance of the sMP state under symmetry
group}\label{appC}

Firstly, we assume that all the operators of the symmetry group $G$
are \textit{unitary}. The CG-coefficients (of the representation
group) are defined as
\begin{eqnarray}\label{CGdef}
&&|m\rangle=\sum_{\alpha,\beta}C^m_{\alpha\beta}|\alpha,\beta\rangle, \nonumber\\
&&|\mathrm{singlet}\rangle=\sum_{\alpha,\beta}B_{\alpha\beta}|\alpha,\beta\rangle
\end{eqnarray}
where $|m\rangle$ belong to nontrivial linear IRs and
$|\mathrm{singlet}\rangle$ is a trivial linear IR, $\alpha,\beta$
are bases of some 2-d projective IR. We will show that the sMP state
given by (\ref{A}) is invariant under the representation group
$R(G)$ (and hence the symmetry group $G$). Suppose that $g$ is a
group element
of $R(G)$, and $u(g)/N(g),M(g)$ is the 
representation matrix for the physical spin/virtual `spins', then
\begin{eqnarray}\label{R}
&&\hat g|m\rangle=u_{m'm}|m'\rangle,\nonumber\\
&&\hat g|\alpha\rangle=N_{\alpha'\alpha}|\alpha'\rangle,\nonumber\\
&&\hat g|\beta\rangle=M_{\beta'\beta}|\beta'\rangle,
\end{eqnarray}
From Eqs.~(\ref{CGdef}) and (\ref{R}), we obtain
\begin{eqnarray}\label{GdD}
\sum_{m'}u_{m'm}C^{m'}=NC^mM^T.
\end{eqnarray}
The complex conjugate of above equation is
\begin{eqnarray}\label{GdD*}
\sum_{m'}u^\dag_{mm'}(C^{m'})^*=N^*(C^m)^*M^\dag.
\end{eqnarray}
Since the representation matrix $u(g)$ ($N(g)$,$M(g)$) is unitary,
so the representation matrix of ${\hat g}^{-1}$ is
$[u(g)]^\dag$($[N(g)]^\dag$,$[M(g)]^\dag$). Replacing $\hat g$ by
${\hat g}^{-1}$ in Eqs.~(\ref{R})-(\ref{GdD*}), we obtain
\begin{eqnarray}\label{GC}
\sum_{m'}u_{mm'}(C^{m'})^*=N^T(C^m)^*M.
\end{eqnarray}
Similar to (\ref{GdD}), we also have
\begin{eqnarray*}
B=NBM^T,
\end{eqnarray*}
or equivalently $B^T=MB^TN^T$. Thus we have
\begin{eqnarray}\label{GB}
M^\dag B^T=B^TN^T.
\end{eqnarray}
From (\ref{GC}) and (\ref{GB}), we have
\begin{eqnarray}\label{symmetry2}
\hat
g\left(\sum_mA^m|m\rangle\right)&=&\sum_{m,m'}u_{m'm}A^m|m'\rangle
\nonumber\\&=&\sum_{m,m'}B^Tu_{m'm}(C^m)^*|m'\rangle\nonumber\\
&=&\sum_{m'}B^TN^T(C^{m'})^*M|m'\rangle\nonumber\\&=&\sum_{m}M^\dag
B^T(C^m)^*M|m\rangle\nonumber\\&=&\sum_{m}M^\dag A^mM|m\rangle.
\end{eqnarray}
Above equation is nothing but (\ref{MAM}), which indicates that the
sMP state constructed by $A^m=B^T(C^m)^*$ is really invariant under
the group $R(G)$ (or equivalently, the symmetry group $G$).

Now we consider the case that some group elements, such as the time
reversal operator $T$, of $G$ are \textit{anti-unitary}. Suppose
that by properly choosing the phases of $|\alpha\rangle$ and
$|\beta\rangle$,  all the CG-coefficients $B$ and $C^m$ are
\textit{real}. In this case, the anti-unitary operators behave as
unitary operators when acting on $A^m$, and (\ref{symmetry2}) also
holds for anti-unitary operators.

To obtain the complete representation of the anti-unitary operators,
we introduce an unitary transformation to the bases $|\alpha\rangle,
|\beta\rangle$ of the virtual `spins' so that $A^m$ transforms into
complex matrix:
\begin{eqnarray}
&&|\alpha\rangle=V_{\alpha'\alpha}|\alpha'\rangle,\nonumber\\
&&|\beta\rangle=U_{\beta'\beta}|\beta'\rangle,
\end{eqnarray}
then
\begin{eqnarray}
|m\rangle&=&C^m_{\alpha\beta}|\alpha\beta\rangle=C^m_{\alpha\beta}V_{\alpha'\alpha}
U_{\beta'\beta}|\alpha'\beta'\rangle\nonumber\\&=&
(VC^mU^T)_{\alpha'\beta'}|\alpha'\beta'\rangle=C^{'m}_{\alpha'\beta'}|\alpha'\beta'\rangle,
\end{eqnarray}
which gives $C^{'m}=VC^mU^T$. Similarly, we have $B'=VBU^T$. Since
$A^m=B^T(C^m)^*$, so we get $A^{'m}=UA^mU^\dag$. When an unitary
operator $\hat g_u$ acts on $A^{'m}|m\rangle$, (\ref{symmetry2})
holds as expected:
\begin{eqnarray*}
\hat g_u(\sum_mA^{'m}|m\rangle)&=&U\hat
g_u(\sum_mA^m|m\rangle)U^\dag\nonumber\\
&=&\sum_mUM^\dag A^m MU^\dag|m\rangle\nonumber\\
&=&\sum_m(M')^\dag A^{'m} M' |m\rangle,
\end{eqnarray*}
where $M'=UMU^\dag$.  Now let us see what happens if an anti-unitary
operator $\hat g_a$ acts on $A^{'m}|m\rangle$:
\begin{eqnarray*}
\hat g_a\left(\sum_mA^{'m}|m\rangle\right)
&=&\hat g_a\left[U(\sum_mA^m|m\rangle)U^\dag\right]\nonumber\\
&=&U^*\hat g_a\left(\sum_mA^m|m\rangle\right)U^T\nonumber\\
&=&\sum_mU^*M^\dag A^{m}MU^T|m\rangle\\
&=&\sum_m(\tilde M')^\dag A^{'m}\tilde M'|m\rangle,
\end{eqnarray*}
where $\tilde M'=UMU^T=U(MK)U^\dag$. Here we have used the factor
that $A^m$ are real matrices. So (\ref{symmetry2}) still holds for
anti-unitary operators, except that the representation matrix
$M(g_a)$ transforms into $\tilde M'(g_a)$ instead of $M'(g_a)$, or
equivalently, $M(g_a)$ is replaced by $M(g_a)K$. Thus for an
anti-unitary operator $\hat g$, we have
\begin{eqnarray}
u(g)K(A^m)&=&KM^\dag A^mMK.
\end{eqnarray}
This result will be used in appendix \ref{appD}. 

Notice that to obtain a sMP state that is invariant under a symmetry
group $G$ containing anti-unitary operators, the only condition we
require is that the CG coefficients $B$ and $C^m$ (for the unitary
projective IRs of $G$) can be transformed into real numbers by
choosing proper phases.

\section{effective operators in the ground state Hilbert
space}\label{appD}

From the projective representation, we can study the effective
operator of a usual operator (which acts on the physical spin
Hilbert space) on the ground state Hilbert space, or equivalently,
the end `spins'. Naturally, the usual operator and its effective
operator should vary in the same way, or respect the same linear
representation, under the group $D_{2h}$. So we will study the
effective operators from the symmetry point of view.

If the spin chain is long enough, the two end `spins' are free
(i.e., the interaction between them are neglectable). So we expect
that the effective operators on the end `spins' are single-body
operators instead of two-body interactions. Notice that the all the
nontrivial projective representations of $D_{2h}$ are 2 dimensional,
we have only three choices of the effective operators, the pauli
matrices. We will study them one by one.

Firstly, we study the $T_0$ phase, which correspond to the
projective IR (-1-1-1). Under symmetry operation $g$, the operator
$\hat O_m$ varies in the following way
\begin{eqnarray}\label{Operator}
M(g)^\dag \hat O_m M(g)=\eta(g)_{mm'}\hat O_{m'},
\end{eqnarray}
where $M(g)$ is the projective IR for the end `spin'. From the
conclusion in appendix \ref{appC} and Tab.~\ref{tab:prjRep}, we get
$M(R_z)=i\sigma_z$, $M(R_x)=i\sigma_x$, $M(T)=i\sigma_yK$. $\eta(g)$
is a linear representation of $D_{2h}$, which equals either $1$ or
$-1$. Actually, $\eta(g)$ is the parity of $\hat O_m$ under $g$. For
instance, $\eta(T)=-1$ means that $\hat O_m$ has odd parity under
time reversal transformation and vice versa. After simple algebra,
we obtain the correspondence in table \ref{tab:effE13}: the
operators in the same column transform in the same way.

\begin{table}[htbp]
\caption{Correspondence between physical operators and effective
operators in $T_0$ phase according to their transformation property
(parities) under $D_{2h}$. } \label{tab:effE13}
\begin{ruledtabular}
\begin{tabular}{c|ccc}
linear IR $\eta(g)$         &     $B_{3u}$      &   $B_{2u}$    & $B_{1u}$\\%
operators $\hat O_m$     &    $\sigma_x$     &  $\sigma_y$   & $\sigma_z$ \\%
physical operators       &    $\tilde S_x$   &  $\tilde S_y$ & $\tilde S_z$\\%
\end{tabular}
\end{ruledtabular}
\end{table}

From above table, we find that $\sigma_m$ and $\tilde S_m$
($m=x,y,z$) have the same symmetry (or the same parity under
symmetry operations), so the former can be considered as the
effective operator of the latter. Since $\sigma_m$ is the spin
operator of the end spins, the system will response to weak external
magnetic field (along any direction) effectively through the end
spins.

However, things are different in $T_x$ phase, which corresponds to
the projective IR (-1-11). From Tab.~\ref{tab:prjRep}, we can
substitute $M(R_z)=i\sigma_z$, $M(R_x)=i\sigma_x$, $M(T)=\sigma_zK$
into (\ref{Operator}) and obtain the results in table
\ref{tab:effE11}.
\begin{table}[htbp]
\caption{Correspondence between physical operators and effective
operators in $T_x$ phase, according to their transformation property
(parities) under $D_{2h}$. } \label{tab:effE11}
\begin{ruledtabular}
\begin{tabular}{c|ccc}
linear IR $\eta(g)$         &     $B_{3u}$      &   $B_{2g}$    & $B_{1g}$\\%
operators $\hat O_m$     &    $\sigma_x$     &  $\sigma_y$   & $\sigma_z$ \\%
physical operators       &    $\tilde S_x$   &$\tilde S_{xz}$& $\tilde S_{xy}$\\%
\end{tabular}
\end{ruledtabular}
\end{table}
Notice that the end `spin' operator $\sigma_y$($\sigma_z$) do not
have the same symmetry with that of $\tilde S_y$($\tilde S_z$)
because they have different time reversal parities. Since there are
no single-body effective operators correspond to $\tilde S_y$ and
$\tilde S_z$, the models in the $T_x$ phase will not response to
weak external magnetic fields along $y$- and $z$- directions.

Similar results can be obtained for $T_y$ and $T_z$ phases and will
not be repeated here.



\bigskip
\noindent

\begin{thebibliography}{99}
\expandafter\ifx\csname natexlab\endcsname\relax\def\natexlab#1{#1}\fi
\expandafter\ifx\csname bibnamefont\endcsname\relax
  \def\bibnamefont#1{#1}\fi
\expandafter\ifx\csname bibfnamefont\endcsname\relax
  \def\bibfnamefont#1{#1}\fi
\expandafter\ifx\csname citenamefont\endcsname\relax
  \def\citenamefont#1{#1}\fi
\expandafter\ifx\csname url\endcsname\relax
  \def\url#1{\texttt{#1}}\fi
\expandafter\ifx\csname urlprefix\endcsname\relax\def\urlprefix{URL }\fi
\providecommand{\bibinfo}[2]{#2}
\providecommand{\eprint}[2][]{\url{#2}}



\bibitem{Wen90} Xiao-Gang Wen, Int. J. Mod. Phys. B4, 239 (1990).

\bibitem{CGW1} Xie Chen, Zheng-Cheng Gu, Xiao-Gang Wen, Phys. Rev. B 82, 155138
(2010).

\bibitem{Haldanephase} F. D. M. Haldane, Phys. Rev. Lett. 50, 1153 (1983), Phys.
Lett. 93,464 (1983); I. Affleck and F. D. M. Haldane, Pyhs. Rev. B
36, 5291 (1987); I. Affleck, J. Phys.: Condens. Matter. l, 3047
(1989).

\bibitem{string} M. den Nijs and K. rommelse, Phys. Rev. B 40, 4709
(1989); T. Kennedy, H. Tasaki, Phys. Rev. B 45, 304 (1992).

\bibitem{GuWen}Zheng-Cheng Gu, Xiao-Gang Wen, Phys.Rev.B 80, 155131
(2009).

\bibitem{CGW2} Xie Chen, Zheng-Cheng Gu, Xiao-Gang Wen, Phys. Rev. B 83, 035107 (2011).

\bibitem{VCL0501} F. Verstraete, J. I. Cirac, J. I. Latorre, E. Rico, and M. M.
Wolf, Phys. Rev. Lett. 94,140601 (2005).

\bibitem{pollmann} F. Pollmann, E. Berg, A. M. Turner, and M. Oshikawa,
arXiv:0909.4059 (2009);Phys. Rev. B 81, 064439 (2010).

\bibitem{entanglspect} H. Li and F. D. M. Haldane, Phys. Rev. Lett. 101, 010504 (2008).

\bibitem[{\citenamefont{Kane and Mele}(2005{\natexlab{a}})}]{KM0501}
\bibinfo{author}{\bibfnamefont{C.}~\bibnamefont{Kane}} \bibnamefont{and}
  \bibinfo{author}{\bibfnamefont{E.}~\bibnamefont{Mele}},
  \bibinfo{journal}{Phys. Rev. Lett.} \textbf{\bibinfo{volume}{95}},
  \bibinfo{pages}{226801} (\bibinfo{year}{2005}{\natexlab{a}}),
  \eprint{cond-mat/0411737}.

\bibitem{BZ}
\bibinfo{author}{\bibfnamefont{B. A.}~\bibnamefont{Bernevig}} \bibnamefont{and}
  \bibinfo{author}{\bibfnamefont{S.-C.}~\bibnamefont{Zhang}},
  \bibinfo{journal}{Phys. Rev. Lett.} \textbf{\bibinfo{volume}{96}},
  \bibinfo{pages}{106802} (\bibinfo{year}{2005}),
  \eprint{cond-mat/0504147}.

\bibitem[{\citenamefont{Kane and Mele}(2005{\natexlab{b}})}]{KM0502}
\bibinfo{author}{\bibfnamefont{C.}~\bibnamefont{Kane}} \bibnamefont{and}
  \bibinfo{author}{\bibfnamefont{E.}~\bibnamefont{Mele}},
  \bibinfo{journal}{Phys. Rev. Lett.} \textbf{\bibinfo{volume}{95}},
  \bibinfo{pages}{146802} (\bibinfo{year}{2005}{\natexlab{b}}),
  \eprint{cond-mat/0506581}.

\bibitem[{\citenamefont{Moore and Balents}(2007)}]{MB0706}
\bibinfo{author}{\bibfnamefont{J.~E.} \bibnamefont{Moore}} \bibnamefont{and}
  \bibinfo{author}{\bibfnamefont{L.}~\bibnamefont{Balents}},
  \bibinfo{journal}{Phys. Rev. B} \textbf{\bibinfo{volume}{75}},
  \bibinfo{pages}{121306} (\bibinfo{year}{2007}), \eprint{cond-mat/0607314}.

\bibitem[{\citenamefont{Fu et~al.}(2007)\citenamefont{Fu, Kane, and
  Mele}}]{FKM0703}
\bibinfo{author}{\bibfnamefont{L.}~\bibnamefont{Fu}},
  \bibinfo{author}{\bibfnamefont{C.}~\bibnamefont{Kane}}, \bibnamefont{and}
  \bibinfo{author}{\bibfnamefont{E.}~\bibnamefont{Mele}},
  \bibinfo{journal}{Phys. Rev. Lett.} \textbf{\bibinfo{volume}{98}},
  \bibinfo{pages}{106803} (\bibinfo{year}{2007}), \eprint{cond-mat/0607699}.

\bibitem[{\citenamefont{Qi et~al.}(2008)\citenamefont{Qi, Hughes, and
  Zhang}}]{QHZ0837}
\bibinfo{author}{\bibfnamefont{X.-L.} \bibnamefont{Qi}},
  \bibinfo{author}{\bibfnamefont{T.}~\bibnamefont{Hughes}}, \bibnamefont{and}
  \bibinfo{author}{\bibfnamefont{S.-C.} \bibnamefont{Zhang}},
Phys. Rev. B {\bf 78}, 195424 (2008),
\eprint{arXiv:0802.3537}.

\bibitem{CGW3} Xie Chen, Zheng-Cheng Gu, Xiao-Gang Wen, arXiv:1103.3323.

\bibitem{e1e2e3} There are two reasons that this phase is `trivial'.
The first one is that it corresponds to the trivial projective
representation of $D_{2h}$, and the second one is that the
entanglement spectrum is not degenerate in the ground state. And
similarly `nontrivial' phases are defined. Actually, there are more
than one `trivial' phases: $e_1\to\infty$, $e_2\to\infty$ and
$e_3\to\infty$ correspond to three `trivial' phases, they are
classified in Ref.~\onlinecite{CGW3} by different phases $\alpha(g)$
given in (\ref{MAM}).


\bibitem{AKLT}I. Affleck, T. Kennedy, E.H. Lieb and H. Tasaki,
Phys. Rev. Lett. 59, 799 (1987); Commun. Math. Phys. 115, 477
(1988).

\bibitem{iMPS} Here a simple matrix product (sMP) state means that it is 
injective,  ergodic, and invariant under the symmetry group. In
literature, such a state is also called a valence bond solid (VBS)
state. Usually, a VBS state is a translational symmetry breaking
state(for spin-1/2 system). To avoid this confusion, we call it sMP
state instead.

\bibitem{Vidal}G. Vidal, Phys. Rev. Lett. 91, 147902 (2003); Phys. Rev. Lett. 93, 040502
(2004); Phys. Rev. Lett. 98, 070201 (2007).

\bibitem{TensRG} H. C. Jiang, Z. Y. Weng, T. Xiang, Phys. Rev. Lett. 101, 090603 (2008).

\bibitem{SU(2)}The three operators $S_x, -S_{xz}, S_{xy}$ still satisfy
angular momentum (or $SU(2)$) algebra, but they do not stand for an
usual spin, because the $y$- and $z$-component operators do not
change their signs under time reversal transformation.

\bibitem{Kittle} Charles Kittel, \textit{Introduction to Solid State
Physics}, Wiley, 8th edition (2004).

\bibitem{Boyal}L. L. Boyle and Kerie F. Green, Mathematical and Physical Sciences, A 288, No. 1351, pp. 237-269
(1978).

\bibitem{Tu}Hong-Hao Tu, Guang-Ming Zhang, Tao Xiang, Zheng-Xin Liu and Tai-Kai Ng, Physical Review B 80, 014401 (2009).

\bibitem{ChenJQ}Jin-Quan Chen, Mei-Juan Gao, and Guang-Qun Ma, Rev. Mod. Phys. 57, 211
(1985); Jin Quan Chen, Jialun Ping, Fan Wang, \textit{Group
Representation Theory For Physicists},  World Scientific Publishing
Company (2002).

\bibitem{D2}Michael M. Wolf, Gerardo Ortiz, Frank Verstraete, and J. Ignacio
Cirac, Phys. Rev. Lett. 97, 110403 (2006).

\end{thebibliography}
\end{document}